\documentclass[reprint,amsmath,amssymb,aps,]{revtex4-2}

\usepackage{float}
\usepackage{caption}
\usepackage{subcaption}
\usepackage{graphicx}
\usepackage{dcolumn}
\usepackage{bm}

\begin{document}

\title{Spin-Symmetry Broken Ground-State of UO$_2$ in DFT+U Approach: The SMC Method}

\author{Mahmoud Payami}%
\email{Email: mpayami@aeoi.org.ir}
\affiliation{School of Physics \& Accelerators, Nuclear Science and Technology Research Institute, AEOI, 
	P.~O.~Box~14395-836, Tehran, Iran
}%

\date{\today}%

\begin{abstract}
It turns out that the ground states of some systems are symmetry-broken states in which some property is not symmetrically distributed. In the case of strongly correlated electron systems, that were studied by the DFT+U method, researchers had shown that the total energy of the system is a multi-minima function of input parameters 
and one has to single out the ground state out of the couples of minimum-energy states. 
However, the methods already introduced to determine these local minimum states were not able to predict all such states which may include the "true" ground state.
In this work, we introduce a new simple and straight-forward method of 
SMC to find the GS as well as the meta-stable states 
of 1k-order anti-ferromagnetic configuration for UO$_2$. 
Using this method, it is shown that the ground state of UO$_2$ system is a spin-symmetry broken state of the electron spin magnetizations of oxygen atoms. Depending on the way we apply the SMC method, we obtain different numbers of meta-stable states, but the same ground states. 
The energetic properties, geometric properties, the electronic density distributions, and the electronic polarization density distributions of the ground state 
and the meta-stable states are shown to be different from each other. These properties also are shown to be sensitive to the magnitude of the initial opposite magnetizations of U1 and U2 
atoms
in the 1k-order anti-ferromagnetic configuration, but the number of meta-stable states as well as the ground-state properties are insensitive to this magnitude. 
Using the PBEsol-GGA approximation for the exchange-correlation we obtain the ground-state properties in excellent agreement with experiments.
	
\end{abstract}


\maketitle

\section{Introduction}\label{sec1}
\begin{figure}
	\centering
	\includegraphics[width=0.8\linewidth]{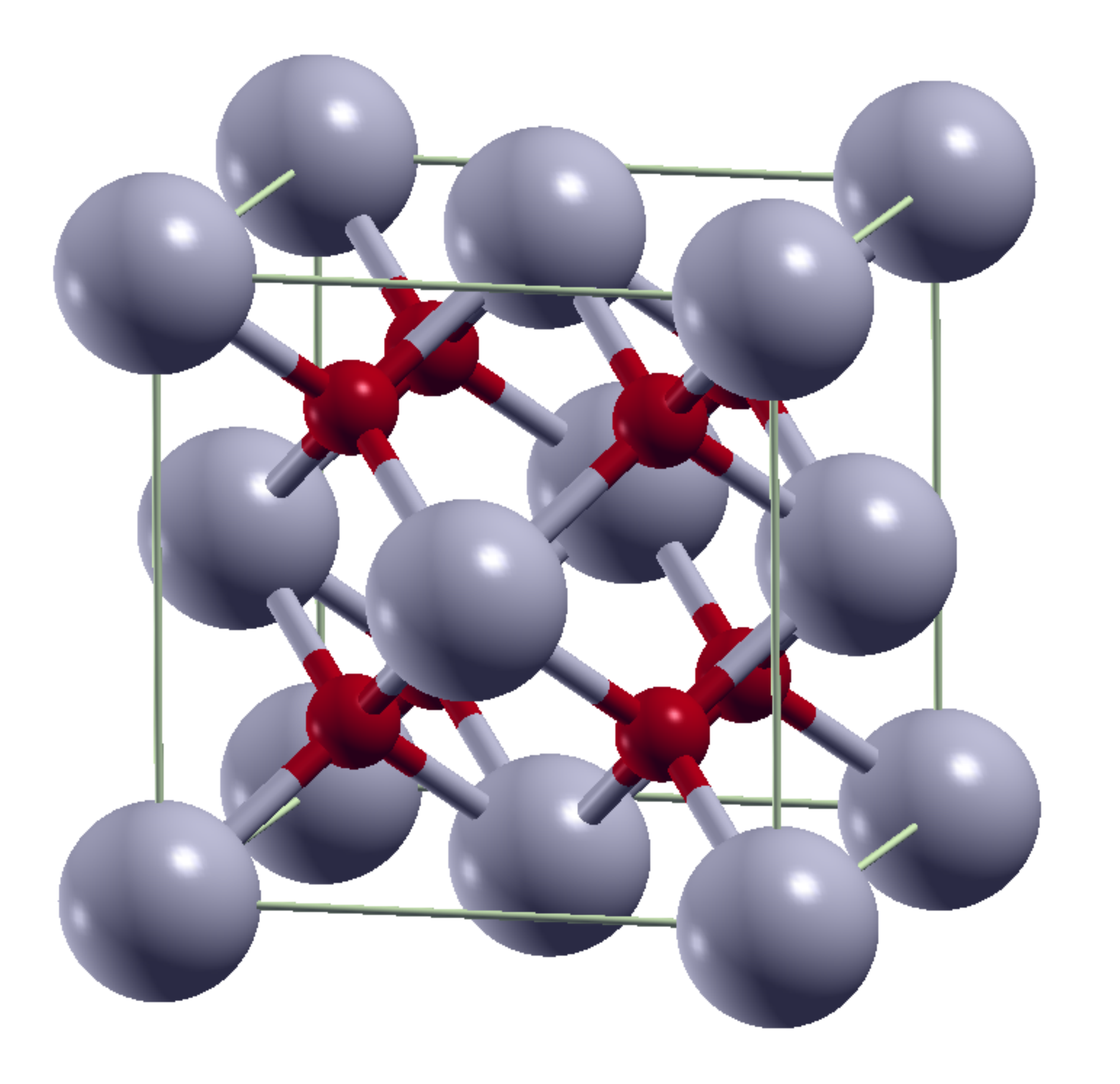}
	\caption{UO$_2$ crystal structure at low temperatures. The U atoms occupy the FCC lattice sites while the oxygen atoms adopt positions with $Pa\bar{3}$ 
		symmetry having an experimental lattice constant equal to 5.47$\AA$.   }
	\label{fig1}
\end{figure}

UO$_2$ is one of the common fuels used in nuclear power reactors. The 
experimental studies had shown that UO$_2$ has an anti-ferromagnetic (AFM) crystal structure with a 3k-order at temperatures less than $30~$K, while it assumes a 
para-magnetic form at higher temperatures\cite{Amoretti,Faber}. In the low-temperature structure, as shown \cite{KOKALJ1999176} in Fig.~\ref{fig1}, the uranium 
atoms occupy the sites of an FCC crystal structure with a lattice constant of $5.47\AA$, and the oxygen atoms occupy positions with $Pa\bar{3}$ 
symmetry\cite{idiri2004behavior}. 

The electronic structure of UO$_2$ has already been investigated by other researchers 
\cite{baer1980electronic,schoenes1978optical,gubanov1977electronic,dudarev1998electronic,schoenes1980electronic,dudarev1997effect,dorado2009dft+,pegg2017dft+,SHEYKHI201893,christian2021interplay}.
It is well-known that the ordinary approximations used in density-functional theory (DFT)\cite{hohenberg1964,kohn1965self} description of the system usually lead 
to incorrect metallic behavior while it is experimentally found to be an insulator, the so-called "Mott insulator". The incorrect metallic prediction arises from 
the usual approximations in which the partially-filled "localized" $5f$ valence electrons in uranium atoms are treated in the same footing as other "delocalized" 
ones in the atom.   
To overcome this problem, one of the ways researchers commonly resort to, is the method of DFT+U 
\cite{cococcioni2005linear,himmetoglu2014hubbard,dorado2009dft+,freyss4} which is also adopted here in our calculations; another method (which is computationally very expensive) is using orbital-dependent 
hybrid functionals for the exchange-correlation (XC) energy functional, which was used by the present author and collaborator \cite{SHEYKHI201893}.  

It has already been noticed that in the DFT+U method, the total energy of the system behaves as a multi-minima function of inputs parameters 
\cite{dorado2009dft+,devey2011first,freyss4,allen2014occupation}. To avoid the non-ground-state energy minima, the so-called "meta-stable" states (MS), 
researchers usually resorted to the methods of occupation-matrix control \cite{dorado2009dft+}, simulated-annealing \cite{geng2010interplay}, and U-ramping 
\cite{meredig2010method}. Each of those methods may help one to find lower-energy states but no guarantee to be the lowest-energy state, i. e., the "true" ground state 
(GS). In this work, we introduce a new simple and straight-forward method of "starting-magnetization control" (SMC) to find the true GS as well as the meta-stable states 
of 1k-order AFM UO$_2$.
Using this method, it is shown that the true GS of UO$_2$ system is a spin-symmetry broken state of the electron spin magnetizations of oxygen atoms. It is worth to mention that the study of symmetry-broken states is one of the hot topics of research in recent decades \cite{PhysRevX.11.041021,perdew2021interpretations,doi:10.1021/acs.jpclett.9b01024,doi:10.1143/JPSJ.67.198,del1991ab,demuth2020evidence,bester2005broken}. 

In SMC method, one scans different starting magnetizations for the two types of oxygen atoms in the interval [-1,+1]  with reasonable 
steps, while the starting magnetizations for the two types of uranium atoms are kept fixed at +0.5 and -0.5. It is shown that using this method in the 
self-consistent (SCF) solution of the spin-polarized Kohn-Sham (KS) equations \cite{kohn1965self}, the system converges to nearest local minimum which is one of the meta-stable states or the 
true GS (From now on for simplicity we omit "true" in the text.).    
Then, one singles out the sub-intervals that lead to the global minimum, i.e., the ground-state, and use them in the further calculations of the ground-state 
properties. 

Examining different XC schemes, we found that the generalized gradient approximation 
(GGA-PBEsol)\cite{PhysRevLett.100.136406,PhysRevLett.102.039902} results in the best agreement with the experimental lattice constant and KS electronic band-gap. 
Therefore, in the calculations of this work we employ GGA-PBEsol.
In the second step, we apply the SMC method to single out the appropriate sub-intervals for the starting magnetization and stick to them for further calculations 
of ground-state properties of UO$_2$. 
In all our calculations, the simplified model of 1k-order AFM configuration for uranium atoms was used. The results for the GS show excellent agreement with the 
experimental lattice constant and electronic band gap.

The organization of this paper is as follows. In Section~\ref{sec2} the computational details are presented; in Section~\ref{sec3} the calculated results are 
presented and discussed; Section~\ref{sec4} concludes this work. 

\section{Computational details}\label{sec2}
All calculations are based on the solution of the KS equations in DFT using the Quantum-ESPRESSO code package \cite{Giannozzi_2009,doi:10.1063/5.0005082}. For the 
atoms U
and O, we have employed the ultra-soft pseudo-potentials (USPP) generated by the {\it atomic} code, using the generation inputs (with small modifications for more 
accurate results) from the {\it pslibrary} \cite{DALCORSO2014337}, at https://github.com/dalcorso/pslibrary.
For the USPP generation, we have used the valence configurations of U($6s^2,\, 6p^6,\, 7s^2,\, 7p^0,\, 6d^1,\, 5f^3 $) and O($2s^2,\, 2p^4 $); and to take into 
account the relativistic effects of the electrons, we have adopted the scalar-relativistic method\cite{koelling1977technique}.

Performing convergency tests, the appropriate kinetic energy cutoffs for the plane-wave expansions
were chosen as 90 and 720~Ry for the wavefunctions and densities, respectively. To avoid the self-consistency problems, we have used the Methfessel-Paxton smearing method \cite{methfessel1989high} 
for the occupations with a width of 0.01~Ry. 
For the Brillouin-zone integrations in geometry optimizations, a $6\times 6\times 6$ grid with a shift were used; while for density-of-states (DOS) calculations, 
we have used a denser grid of $8\times 8\times 8$ in reciprocal space and "tetrahedron" method \cite{PhysRevB.49.16223} for the occupations. In DFT+U 
calculations, we
have used the optimum value of 4.0 eV for Hubbard-U parameter, consistent with the values determined by other
works \cite{yamazaki1991systematic,kotani1992systematic}. All geometries were fully optimized for total pressures on unit cells to within 0.5 kbar, and forces on 
atoms to within 10$^{-6}$~Ry/a.u.

To apply the SMC method, we have first considered one and the same degrees of freedom for the starting magnetization of all oxygen atoms in the unit cell and 
after optimization of the structures, obtained sixteen different local-minimum energy states (some of them doubly-degenerate) including the GS. However, the spin-alignments of uranium atoms in the 
1k-order AFM configuration imply that the oxygen atoms in the planes near to the planes of uranium atoms with different spin-alignments may behave independently. 
We have therefore released the constraint of all-equivalent oxygen atoms, and treated the oxygen atoms near to inequivalent uranium atoms as inequivalent 
ones.     
In this way, we have tried different starting magnetizations for the two types of oxygen atoms separately, while the starting magnetizations for the two types of 
uranium atoms were kept fixed at +0.5 and -0.5. Consequently, the optimization of the structures leads to more meta-stable states compared to all-equivalent oxygen 
atoms treatment. We have also shown that applying the method of occupation matrix control on top of the SMC-determined states may lead to some other meta-stable 
states predicted when the SMC applied with two inequivalent oxygen atoms. 

\section{Results and Discussions}\label{sec3}
To make our calculation results in most agreement with experiments, we first choose the best XC which results in lattice constant and electronic band-gap closest 
to the experimental one. 
Due 
to modeling the low-temperature system with a 1k-order AFM, the lattice constant along $z$ direction becomes slightly different from that in perpendicular 
direction. Using DFT+U and choosing Hubbard parameter $U=4.0~eV$, we have obtained the best values for the equilibrium lattice constants and the electronic 
band-gap using the PBEsol \cite{PhysRevLett.100.136406,PhysRevLett.102.039902}, and the result is shown in Table~ \ref{tab1}. In all subsequent calculations we 
use this XC functional.  

\begin{table}[b]
	\caption{\label{tab1}%
		Equilibrium lattice constants in $\AA$, total and absolute magnetizations in Bohr-magneton per unit formula, and the electronic band gap in eV, using the 
		PBEsol for the XC.}
	\begin{ruledtabular}
		\begin{tabular}{lcccr}
		  $a$ ($c$)   &  $a$ (Exp.)  & Tot. mag.  & Abs. mag. &  $E_g$  \\ \colrule 
       5.5086 (5.4796)& 5.470        & 0.00   & 2.165   & 2.10   \\ 
		\end{tabular}
	\end{ruledtabular}
\end{table}

\subsection{Determining meta-stable states using the SMC method}
\subsubsection{All-equivalent O-atoms}
\begin{figure}
	\centering
    \begin{subfigure}{.23\textwidth}
	\includegraphics[width=\textwidth]{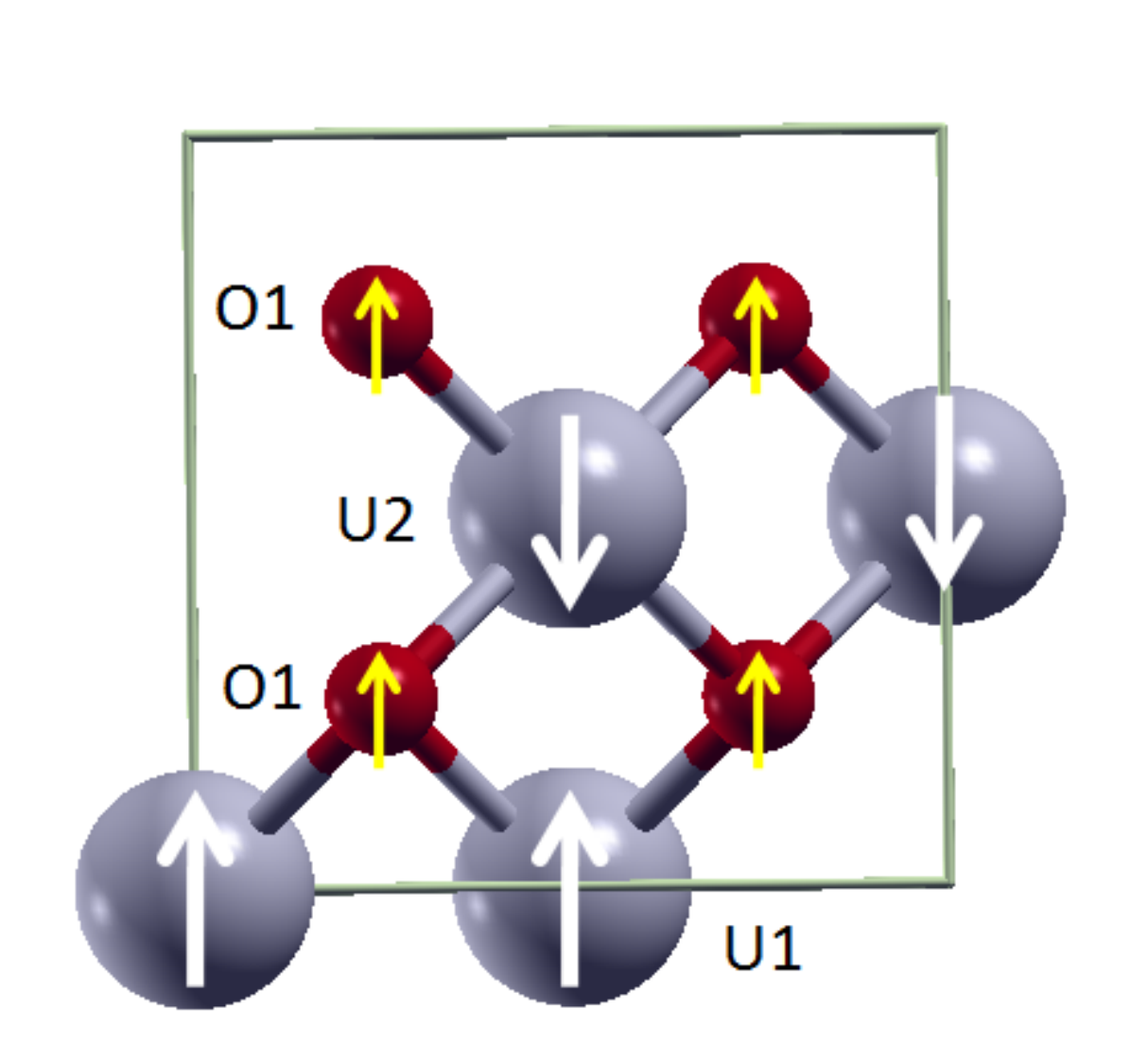}
	\end{subfigure}
    \begin{subfigure}{.23\textwidth}
	\includegraphics[width=\textwidth]{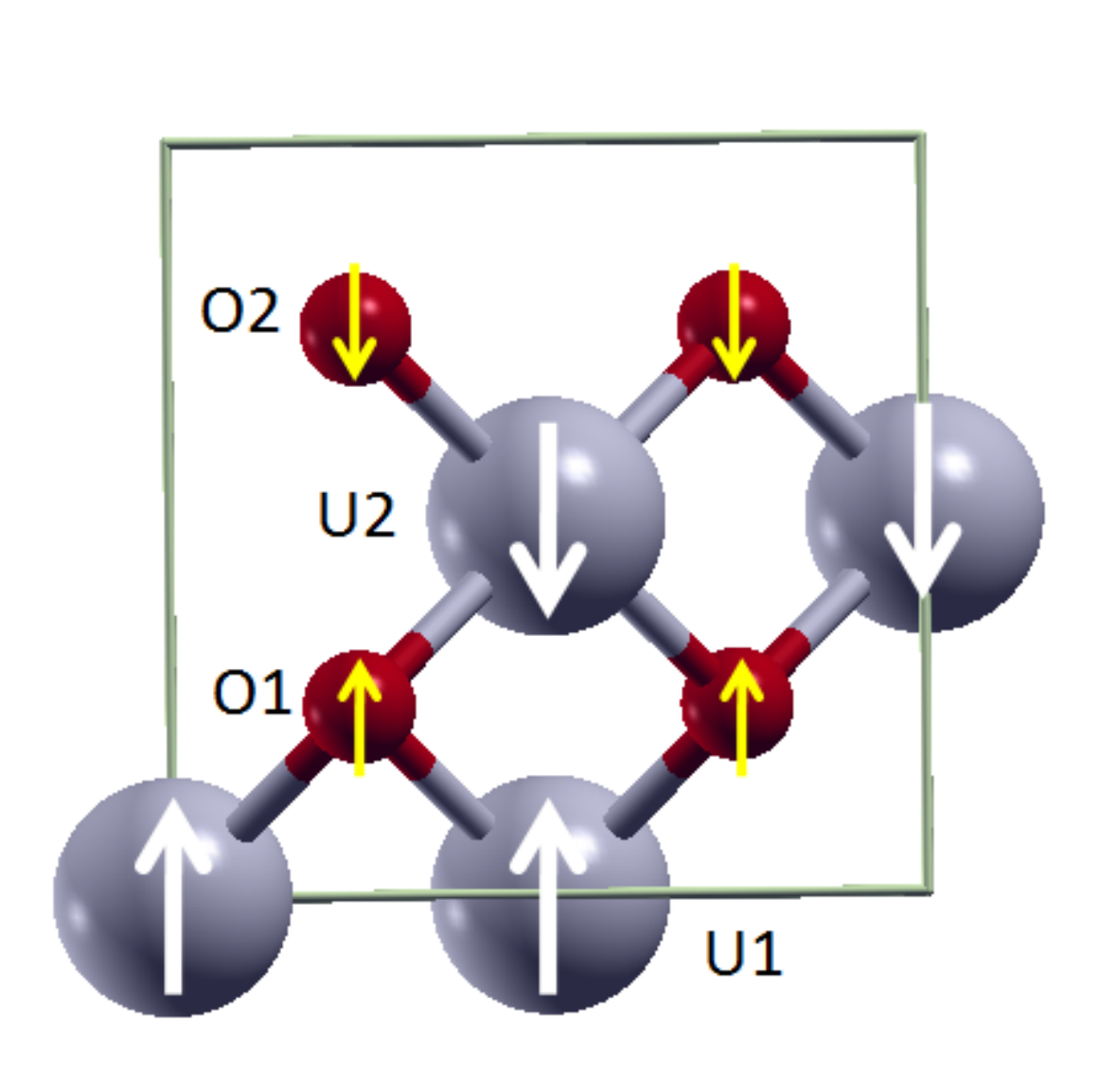}
	\end{subfigure}\\
	\caption{All-equivalent oxygen atoms scheme (left) and two-inequivalent oxygen atoms scheme (right). In the all-equivalent oxygen atoms model, all oxygen 
	atoms are treated as the same type, O1, and assume the same starting magnetizations within the SMC method; while in two-inequivalent oxygen atoms model,
	the O1 and O2 oxygen atoms are treated as different types and assume independent values for the starting magnetizations in the SMC method.}
	\label{fig2}
\end{figure}

As was mentioned earlier, we first consider three types of atoms in the cell, as shown in Fig.~\ref{fig2}. In this model, all oxygen atoms are treated as the same 
type and assume the same starting magnetizations within the SMC method.
In this model, the spin-polarized KS equations were solved for all possible starting magnetizations for the O atoms, while those for U atoms were kept fixed at 
+0.5 and -0.5 Bohr-magnetons per atom. It should be mentioned that the magnetization for an atom with respective $N_\uparrow$ and $N_\downarrow$ spin-up and 
spin-down valence electrons is given by $\zeta=(N_\uparrow-N_\downarrow)/(N_\uparrow+N_\downarrow)$ which varies between -1 and +1. The geometries were fully optimized for each value of the starting magnetization of O1 atom. The results show seven different classes of energetic and structural properties which are summarized in Table~\ref{tab2}. 

\begin{table}[b]
	\caption{\label{tab2}%
		GS and meta-stable states' properties in the simplified all-equivalent oxygen atom model. The energies are in Ry/(unit formula) and are compared to the GS. Equilibrium lattice constants are in $\AA$, total and absolute magnetizations are in Bohr-magneton/(unit formula). GS, M2, and M3 are doubly degenerate states.}
	\begin{ruledtabular}
		\begin{tabular}{lcccr}
			State &   $\Delta E$    &  $a$ ($c$)      &  Tot. mag.  & Abs. mag. \\ \colrule 
			GS    &   0.0000        &  5.5086 (5.4796)&  $-0.00$    & 2.165     \\ 
			GS'   &   0.0000        &  5.5086 (5.4796)&  $+0.00$    & 2.165     \\
			M1    &   0.0027        &  5.5219 (5.4562)&  $\pm 0.00$ & 2.155   \\
			M2    &   0.0118        &  5.5299 (5.4396)&  $-0.00$    & 2.165   \\
			M2'   &   0.0118        &  5.5299 (5.4396)&  $+0.00$    & 2.165   \\
			M3    &   0.0588        &  5.4690 (5.5103)&  $-0.04$    & 2.240   \\
			M3'   &   0.0588        &  5.4690 (5.5103)&  $+0.04$    & 2.240   \\			
		\end{tabular}
	\end{ruledtabular} 
\end{table}

The calculated total magnetizations $M_{tot}= \int_{cell}{ \left( n_{\uparrow}-n_{\downarrow} \right) d^3r}$ 
and absolute magnetizations $M_{abs} = \int_{cell}{ \left| n_{\uparrow}-n_{\downarrow} \right| d^3 r}$ show that the GS as well as most of meta-stable states have 
zero total magnetizations. The $n_\uparrow$ and $n_\downarrow$ are the spin-up and spin-down electron densities, respectively.
To continue calculations for the GS, one simply uses the initial magnetizations giving rise to the GS energy.

\begin{figure}
	\centering
	\includegraphics[width=0.8\linewidth]{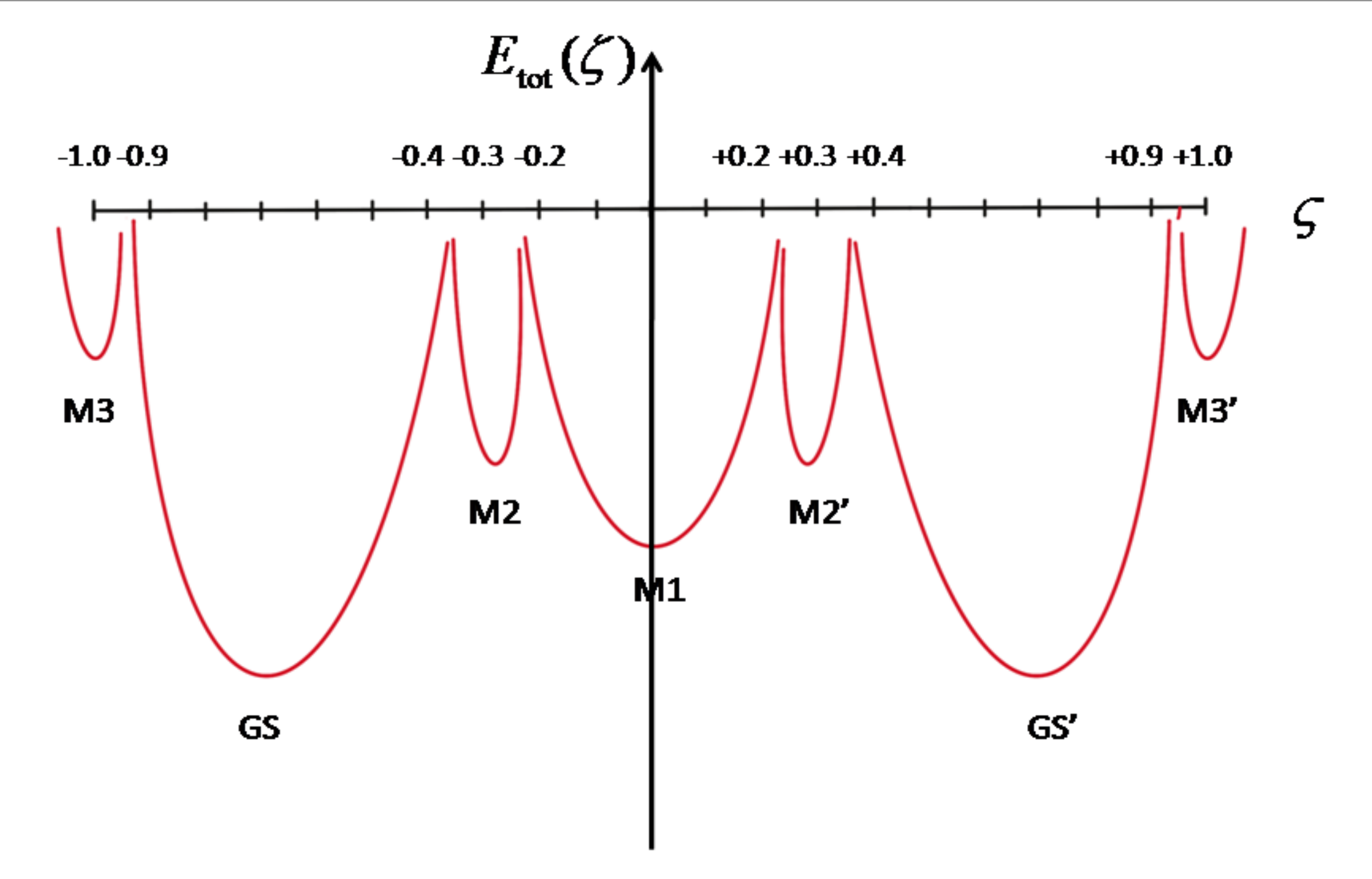}
	\caption{Schematic plot of local minima and their corresponding starting magnetizations. The depths of the minima (in arbitrary units) are consistent with 
	their energy values. As is seen, the GS and GS' cover the largest interval, but away from the zero starting magnetization.   }
	\label{fig3}
\end{figure}

To make the situation more clear, we have schematically shown in Fig.~\ref{fig3} how each starting magnetization leads to the corresponding local minimum. As
is shown in Fig.~\ref{fig3}, considering small values for the starting magnetization, in order to break the symmetry between spin-up and spin-down polarization in a spin-polarized calculation with DFT+U, does not necessarily lead to the GS. 

Now, applying the method of occupation-matrix control\cite{dorado2009dft+} on the determined meta-stable states show that for starting magnetizations $\pm 1.0$ we obtain an extra meta-stable state with $\Delta E=0.0289$, zero total magnetization, 2.24 for the absolute magnetization, and $a(c)=5.4803(5.5020)$. As will be shown in the next subsection, this new state is straightforwardly predicted when we increase the oxygen-atom degrees of freedom.  

\subsubsection{Two inequivalent O-atoms}
In this generalization, we distinguish four types of atoms in the unit cell, as shown in Fig.~\ref{fig2}: U1(0.00), U2(0.50), O1(0.25), and O2(0.75), where the values within parentheses specify the $z$-components of atomic positions in units of $a=5.47\AA$, before geometry optimization. 

To apply the SMC method to find the meta-stable states 
as well as the GS, we keep fixed the starting magnetizations of U1 and U2 atoms at +0.5 and -0.5, respectively; and change the values for O1 and O2 atoms 
independently, in the interval [-1,+1] with steps of 0.1.
Then, for each pair of initial magnetizations $\zeta_1$ and $\zeta_2$, we perform a structural optimization as explained in the previous subsection. 
In this case, we obtain the GS and 16 meta-stable states with energies within 0.0588~Ry/(unit formula) above the GS. Our results show that the number of 
meta-stable states is about two times the number of states reported in previous other works \cite{dorado2009dft+,christian2021interplay}. As shown in 
Fig.~\ref{fig3} corresponding to the simplified model, some meta-stable states cover larger starting-magnetization interval than others. For example, the GS 
covers the interval $\zeta\in [-0.9,-0.4]$ 
and the GS' covers $\zeta\in [+0.4,+0.9]$, so that the total occurrences for the lowest-energy state is 12 out of total 21 different $\zeta$ values in the 
interval $[-1.0,+1.0]$. We therefore include, in the generalized case, the occurrences of the states as well. Here we have named the meta-stable states as "MS" to distinguish them from those of simplified model, "M". As is seen, the states obtained in the simplified model are also included in the generalized case but with different naming. 

\begin{table}[b]
	\caption{\label{tab3}%
		GS and meta-stable states' properties in the two-inequivalent O1 and O2 model. The energies are in Ry/(unit formula) and are compared to the GS. Equilibrium lattice constants are in $\AA$, total and absolute magnetizations are in Bohr-magneton/(unit formula). }
	\begin{ruledtabular}
		\begin{tabular}{lccccr}
			State &   $\Delta E$    &  $a$ ($c$)      &  Tot. mag.  & Abs. mag. & Occ. \\ \colrule 
			GS    &   0.00000       &  5.5086 (5.4796)&  0.00    & 2.165 &  187    \\ 
			MS1   &   0.00275       &  5.5219 (5.4562)&  0.00    & 2.155 &  117  \\
            MS2   &   0.00539       &  5.5040 (5.4835)&  0.00    & 2.155 &   6   \\
           	MS3   &   0.01184       &  5.5299 (5.4396)&  0.00    & 2.165 &   53  \\
           	MS4   &   0.02889       &  5.4803 (5.5020)&  0.00    & 2.240 &   5   \\
           	MS5   &   0.02893       &  5.4855 (5.4861)&  0.00    & 2.240 &   3   \\
           	MS6   &   0.03008       &  5.5054 (5.4740)&  0.00    & 2.180 &   4   \\
           	MS7   &   0.03040       &  5.4939 (5.4738)&  0.00    & 2.240 &   17  \\
           	MS8   &   0.03046       &  5.4848 (5.4932)&  0.00    & 2.245 &   7   \\
           	MS9   &   0.03176       &  5.4851 (5.4874)&  0.00    & 2.235 &   1   \\
           	MS10  &   0.03186       &  5.4848 (5.4932)&  0.00    & 2.205 &   1   \\
           	MS11  &   0.03192       &  5.4842 (5.4637)&  0.00    & 2.230 &   1   \\
           	MS12  &   0.05764       &  5.4657 (5.4895)&  0.00    & 2.300 &   25  \\
           	MS13  &   0.05767       &  5.4746 (5.4749)&  0.00    & 2.300 &   1   \\
           	MS14  &   0.05774       &  5.4701 (5.4829)&  $-0.025$ & 2.300 &   2   \\
           	MS15  &   0.05790       &  5.4647 (5.4433)&  0.00    & 2.290 &   2   \\
           	MS16  &   0.05879       &  5.4690 (5.5103)&$\pm 0.04$& 2.240 &   2   \\
       \end{tabular}
	\end{ruledtabular} 
\end{table}

In Table~\ref{tab3}, the results for local-minima states of the generalized case are summarized. In this generalized case of two inequivalent O1 and O2 atoms, we 
have totally $21\times 21=441$ different pairs of ($\zeta_1 , \zeta_2$) values. 9 cases did not 
converge, and so the sum of occurrences in Table~\ref{tab3} amounts to 441 cases. As is seen, the meta-stable states of M1, M2, and M3 in the simplified model 
are reappeared as MS1, MS3, and MS16 in the generalized model. Here, also as in the simplified model, the small values for both starting magnetizations do not converge to the GS. The meta-stable state obtained from the occupation-matrix control method, discussed at the end of previous subsection, is specified as MS4. 

\subsection{Properties of the GS and MS's}
In this section, we compare the electronic structure properties of the GS with those of M1, M2, and M3, as defined in the simplified model.

\subsubsection{Density of states (DOS)}  

\begin{figure*} 
	\centering
	\begin{subfigure}{.24\textwidth}
				\includegraphics[width=\textwidth]{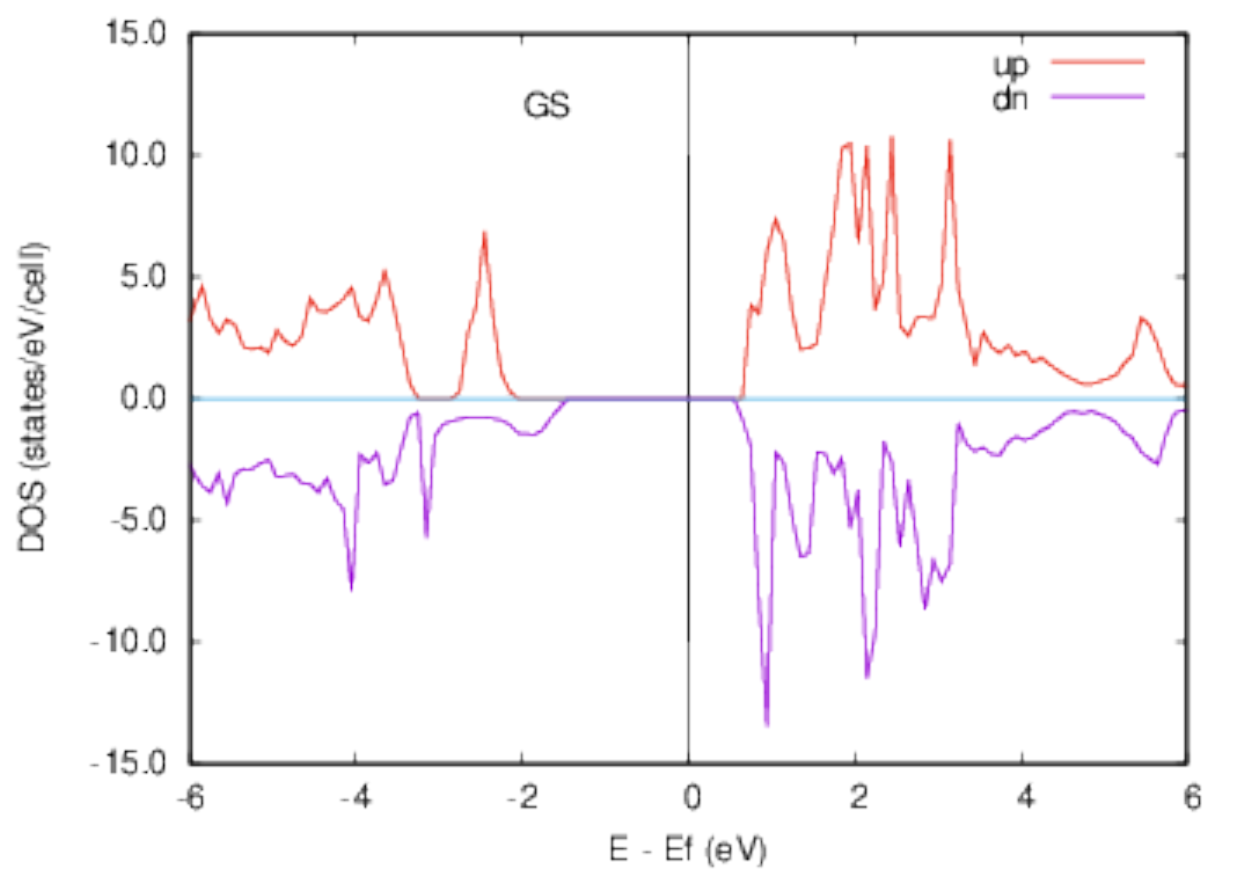}
	\end{subfigure} 
	\begin{subfigure}{.24\textwidth}
			\includegraphics[width=\textwidth]{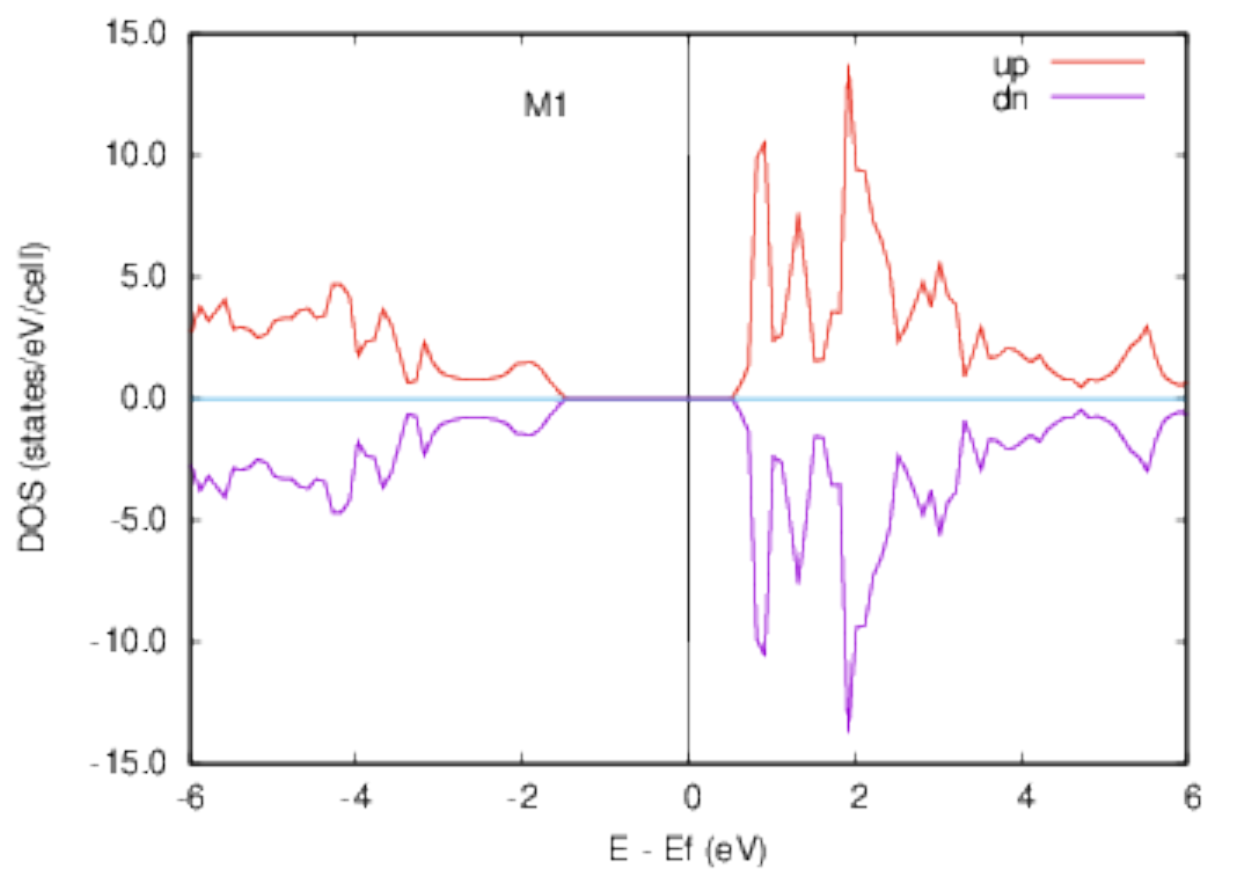}
	\end{subfigure} 
	\begin{subfigure}{.24\textwidth}
			\includegraphics[width=\textwidth]{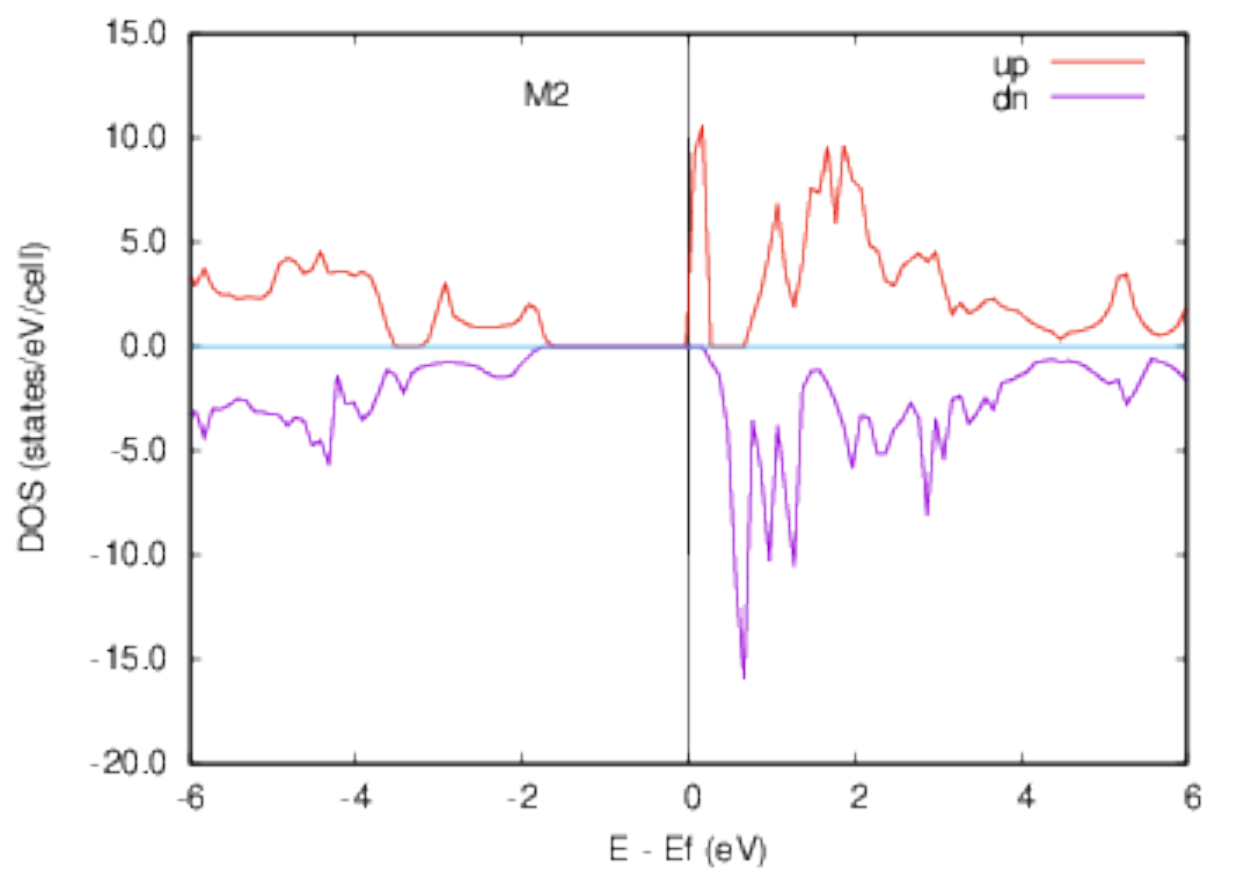}
	\end{subfigure} 
	\begin{subfigure}{.24\textwidth}
			\includegraphics[width=\textwidth]{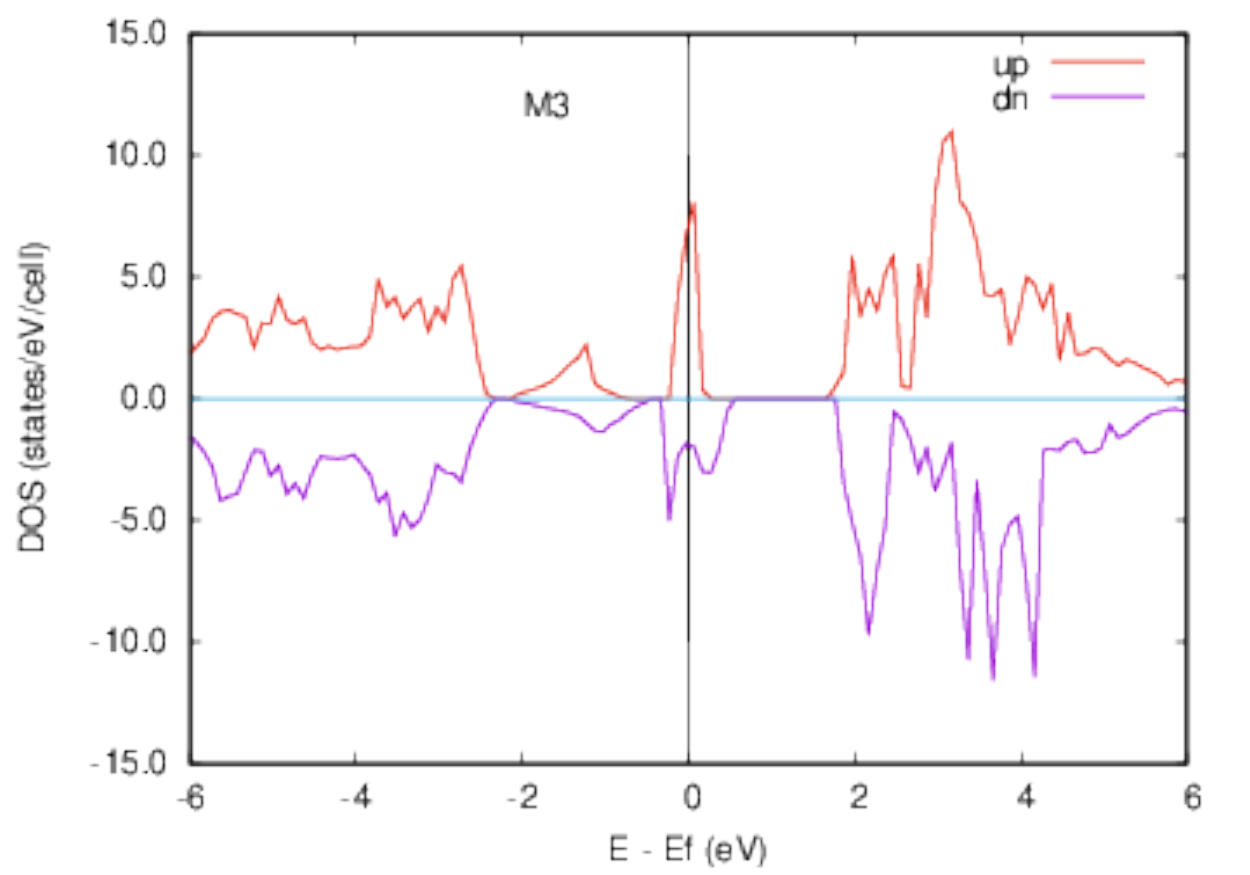}
	\end{subfigure} \\
	\caption{Spin-up and spin-down density of states for the GS and meta-stable states M1, M2, M3. The first three ones GS, M1, M2 are insulators, while M3
		shows metallic behavior.}
	\label{fig4}
\end{figure*}

In Fig.~\ref{fig4}, the spin-up and spin-down density of states for the ground state GS and meta-stable states M1, M2, and M3 are compared. As is seen, the first 
three states GS, M1, and M2 are insulators, with electronic gaps of 2.10, 2.10, and 1.70 eV, respectively; while the meta-stable state M3 shows a narrow-band 
metallic behavior. The small gap in the valence band around 3.0 eV below the Fermi level for the spin-up GS comes from the uranium $5f$ orbitals, as will be seen 
from the projected density of states (PDOS) in the following. A similar small gap is also observed in the conduction band of meta-stable state M2 around 0.5 eV 
above the Fermi level for spin-up.

\begin{figure} 
	\centering
	\begin{subfigure}{.49\textwidth}
		\includegraphics[width=\textwidth]{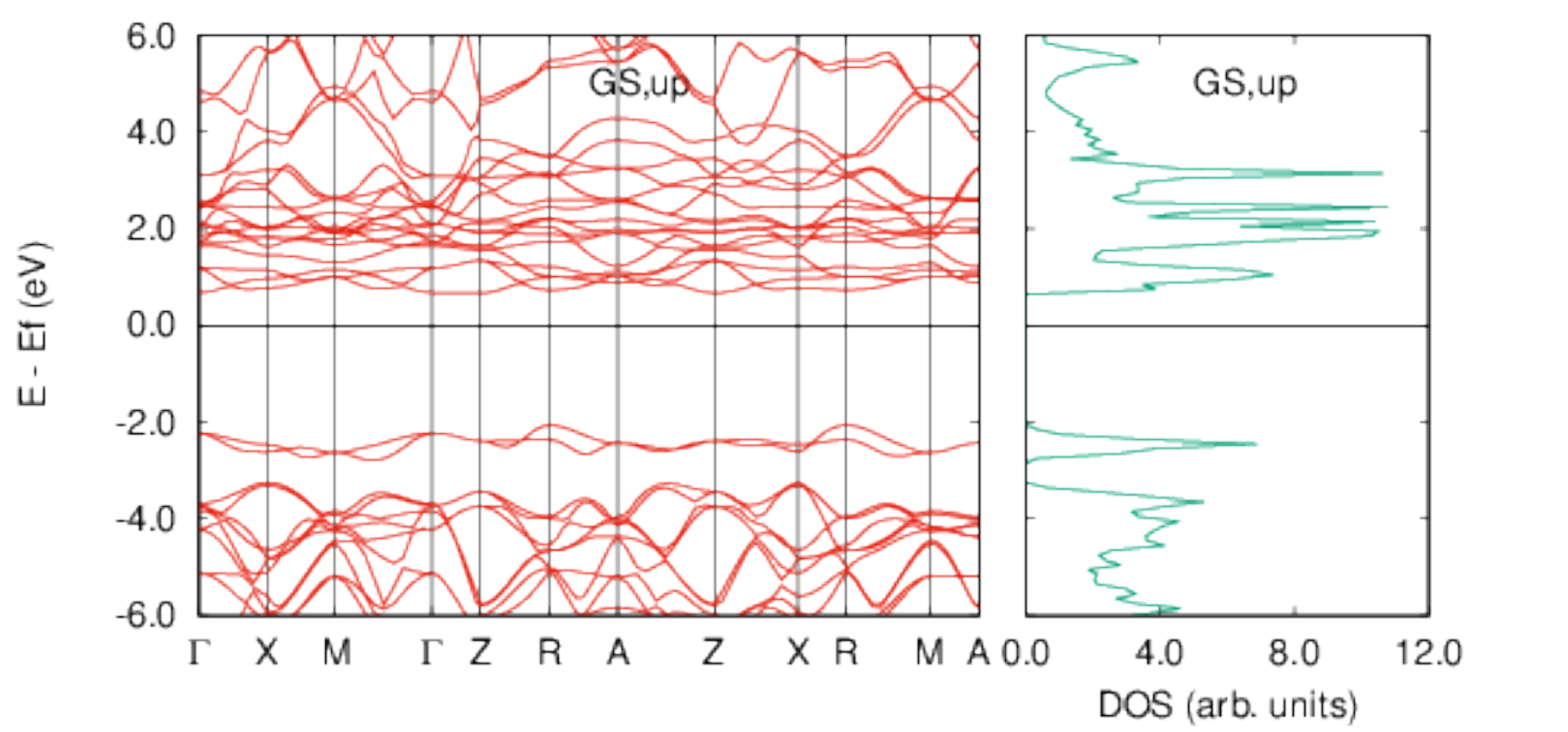}
	\end{subfigure} \\
	\begin{subfigure}{.49\textwidth}
		\includegraphics[width=\textwidth]{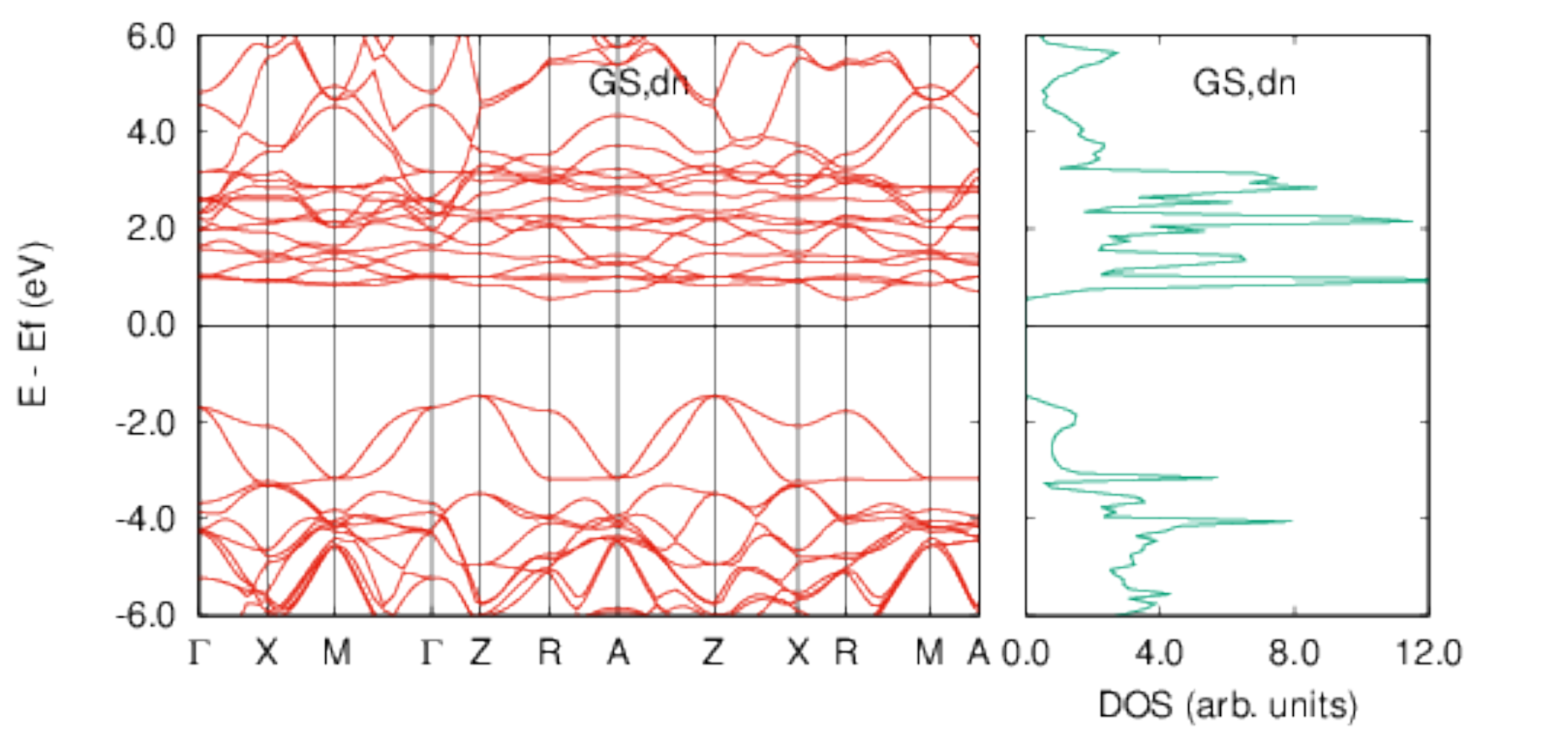}
	\end{subfigure} \\
	\caption{Spin-up and spin-down density of states and their corresponding band structures for the GS.}
	\label{fig5}
\end{figure}

\begin{figure}
	\centering
	\includegraphics[width=\linewidth]{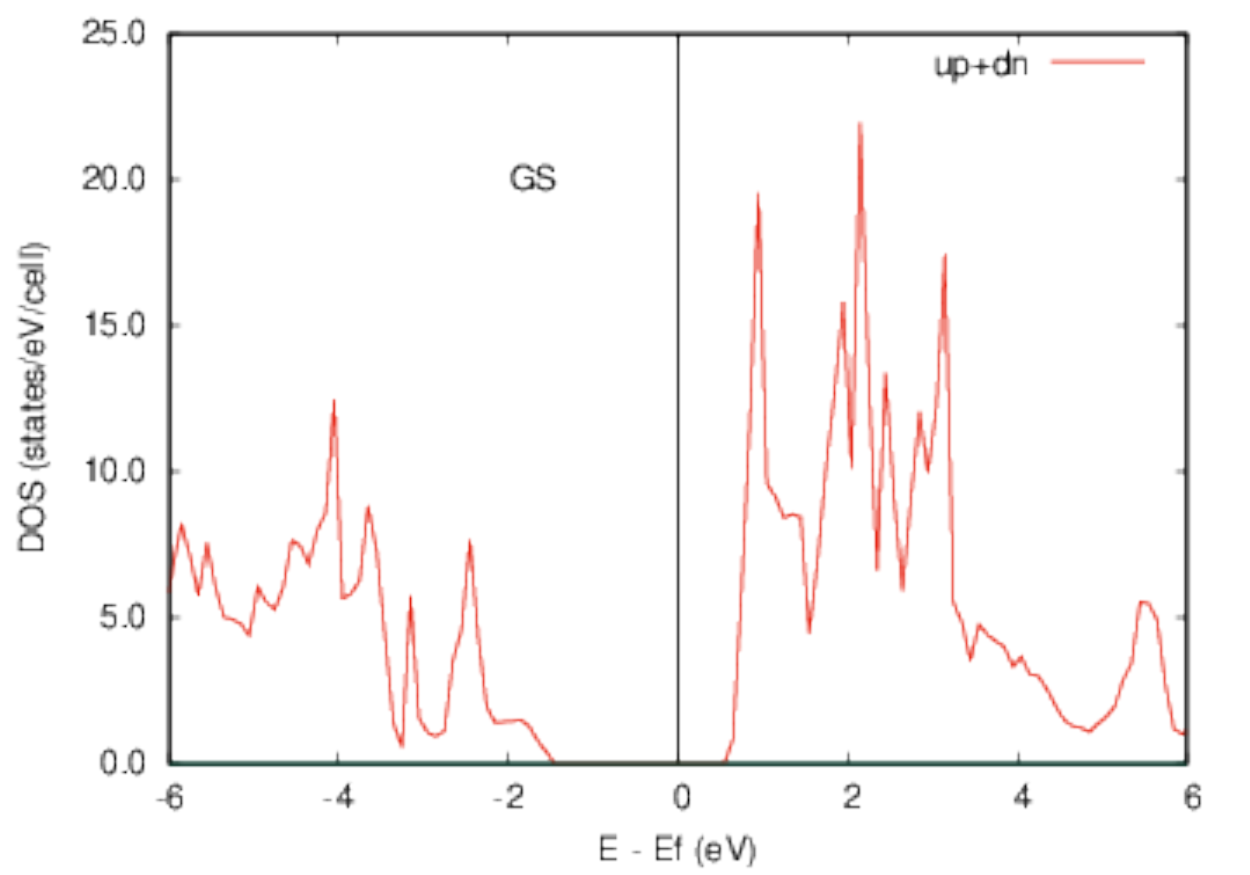}
	\caption{Total DOS for the GS. The small gap of spin-up as shown in Fig.~\ref{fig6} is disappeared. The electronic band gap is determined from the total-DOS to be 2.10 eV.}
	\label{fig6}
\end{figure}

In Fig.~\ref{fig5}, the spin-up and spin-down electronic band structures with their corresponding density of states for the GS are shown. To calculate the band 
structures, we have used the appropriate k-point pathway\cite{HINUMA2017140,togo2018textttspglib} of $\Gamma - X - M - \Gamma - Z - R - A - Z - X - R - M - A$.
As was discussed earlier, the small band-gap in the spin-up is also present in the corresponding band structure. The narrow valence band of spin-up around 2.0 eV 
below the Fermi level originates from the uranium $5f$ orbitals. The valence band-edge here, however, is determined by the spin-down states and the small gap is 
not present in the total density of states plot, as shown in Fig.~\ref{fig6}. The electronic band gap is determined by the total-DOS which is 2.10 eV for the GS.       

\subsubsection{Projected density of states (PDOS)}
To analyze the contributions of each valence atomic orbital to the density of states, we use the projections of wavefunctions over atomic orbitals, and then 
calculate DOS projected onto atomic orbitals, named PDOS. Here, we have plotted the contributions of five valence atomic orbitals $6s$, $6p$, $7s$, $6d$, $5f$ of 
U-atoms 
and the two valence atomic orbitals $2s$, $2p$ of O-atoms for all states GS, M1, M2, and M3.

\begin{figure*}
	\centering
	\begin{subfigure}{0.24\textwidth}
		\includegraphics[width=\linewidth]{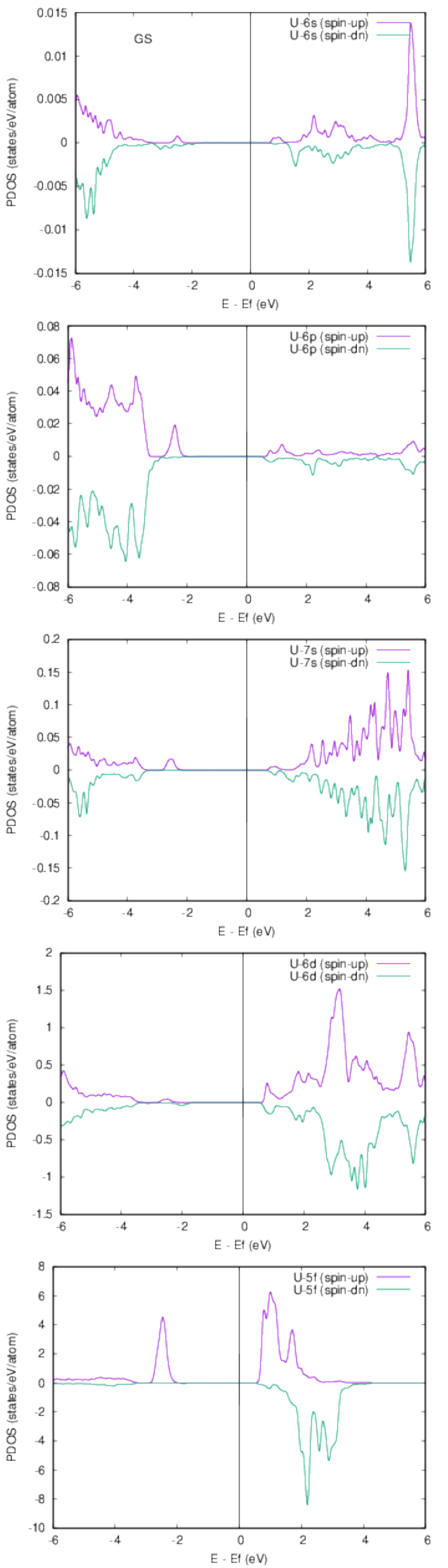}
	\end{subfigure}
	\begin{subfigure}{0.24\textwidth}
		\includegraphics[width=\linewidth]{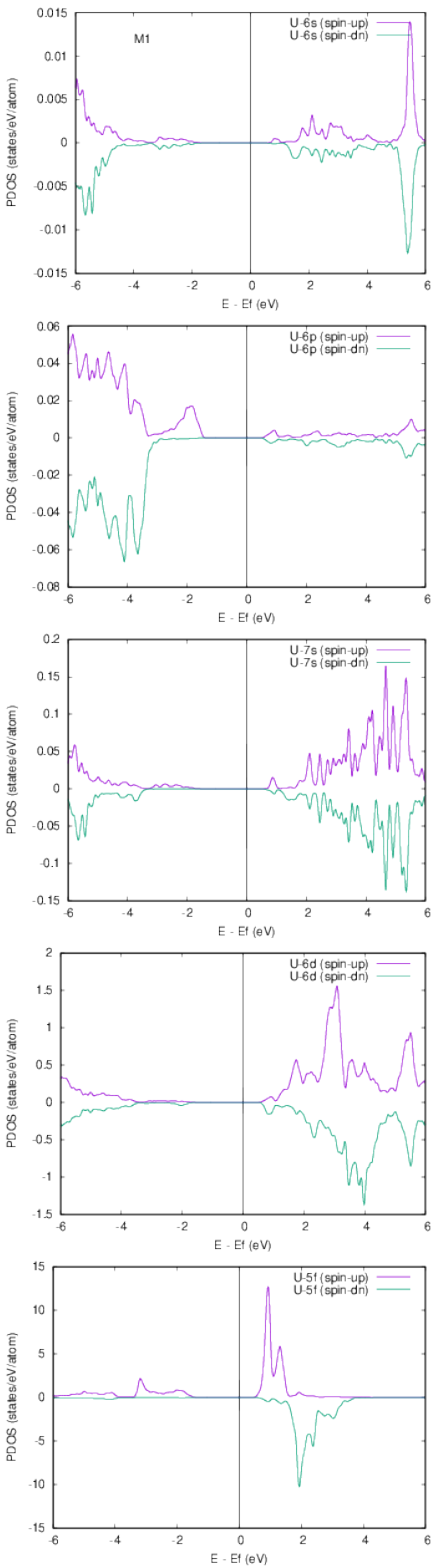}
	\end{subfigure}
	\begin{subfigure}{0.24\textwidth}
		\includegraphics[width=\linewidth]{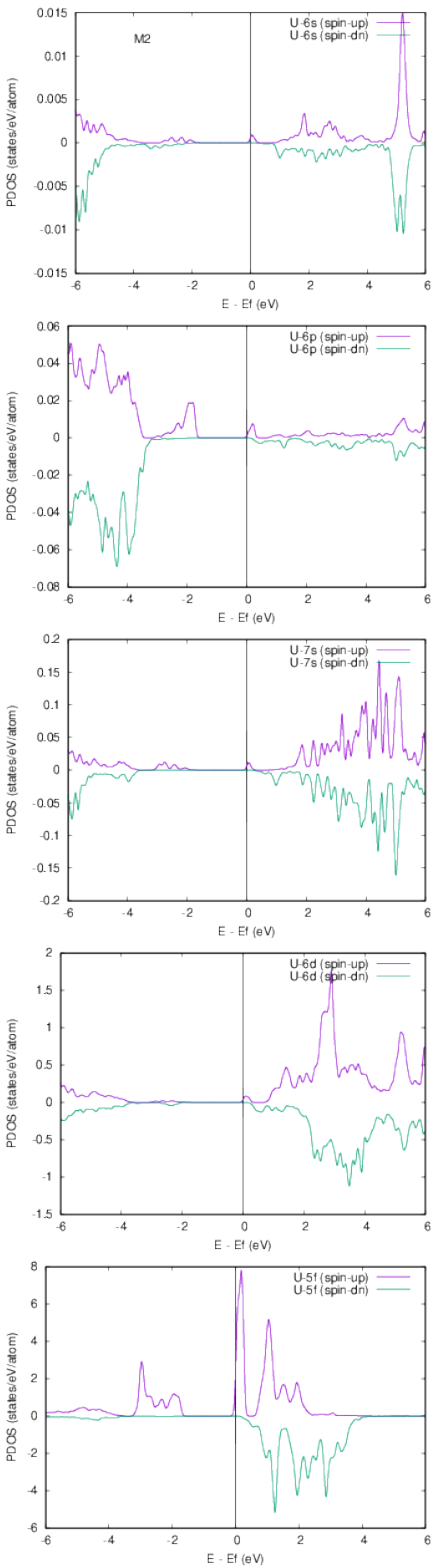}
	\end{subfigure}
	\begin{subfigure}{0.24\textwidth}
		\includegraphics[width=\linewidth]{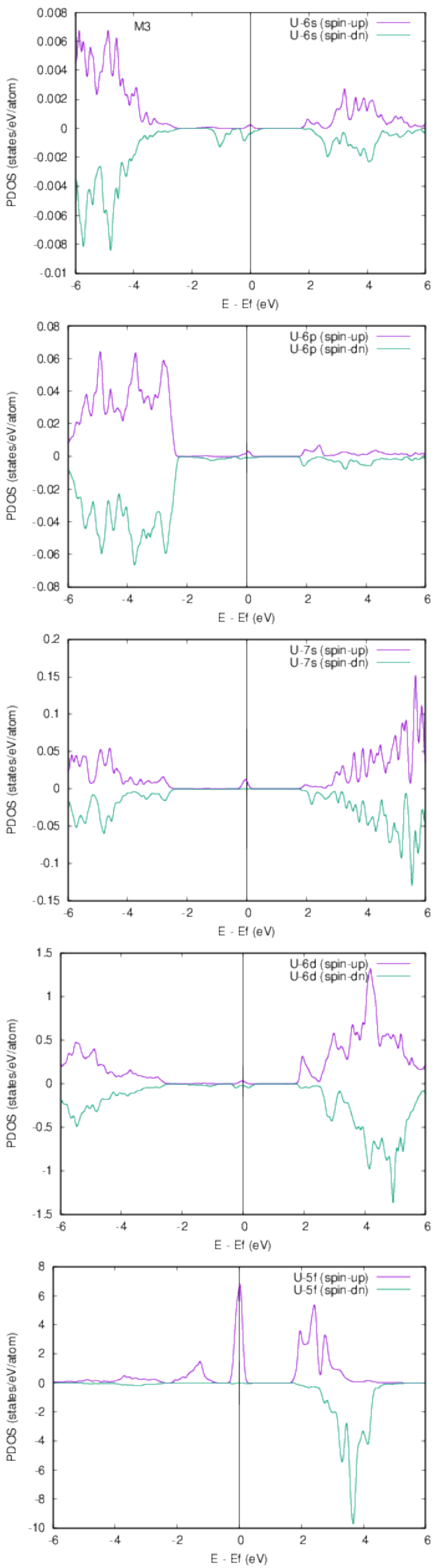}
	\end{subfigure}\\
    \caption{Projected density of states onto five atomic valence atomic orbitals $6s$, $6p$, $7s$, $6d$, $5f$ of 
    	U-atoms. From different ranges of values of different orbitals' PDOS, it is seen that the $5f$ orbitals have the strongest contribution in the density of 
    	states of both valence and conduction bands in all GS, M1, M2, and M3. None of the atomic orbitals have contributions to DOS of GS, M1, and M2 states at 
    	the Fermi level, and therefore, these three states show insulating behaviors. However, in M3 all atomic orbitals have contributions to DOS at Fermi level, 
    	and so this state has metallic behavior. }
	\label{fig7}
\end{figure*}

It is seen from Fig.~\ref{fig7} that the $5f$ orbitals have the strongest contribution in the density of states of both valence and conduction bands in all GS, M1,
M2, and M3. This fact is clear from the different ranges of values of PDOS for different orbitals.
It is seen that the spin-down contributions of $5f$ in the valence band are negligible compared to those of spin-up for all states. 
The contributions of $6d$ are almost similar for the spin-up and spin-down. The strength of $6d$ orbitals are smaller than those of $5f$, but of the same order. On the other hand, the strength of $7s$ orbitals are one order of magnitude smaller than those of $5f$ and $6d$. The strengths of $6p$ and $6s$ contributions are two and three orders of magnitudes smaller than $5f$, respectively. 
None of the atomic orbitals show any contributions to DOS of GS, M1, and M2 states at 
the Fermi level, and therefore, these three states have insulating behaviors. However, it is seen from Fig.~\ref{fig7} that in M3 all atomic orbitals have 
contributions to DOS at Fermi level, and so this state has metallic behavior.

\begin{figure*}
	\centering
	\begin{subfigure}{0.24\textwidth}
		\includegraphics[width=\linewidth]{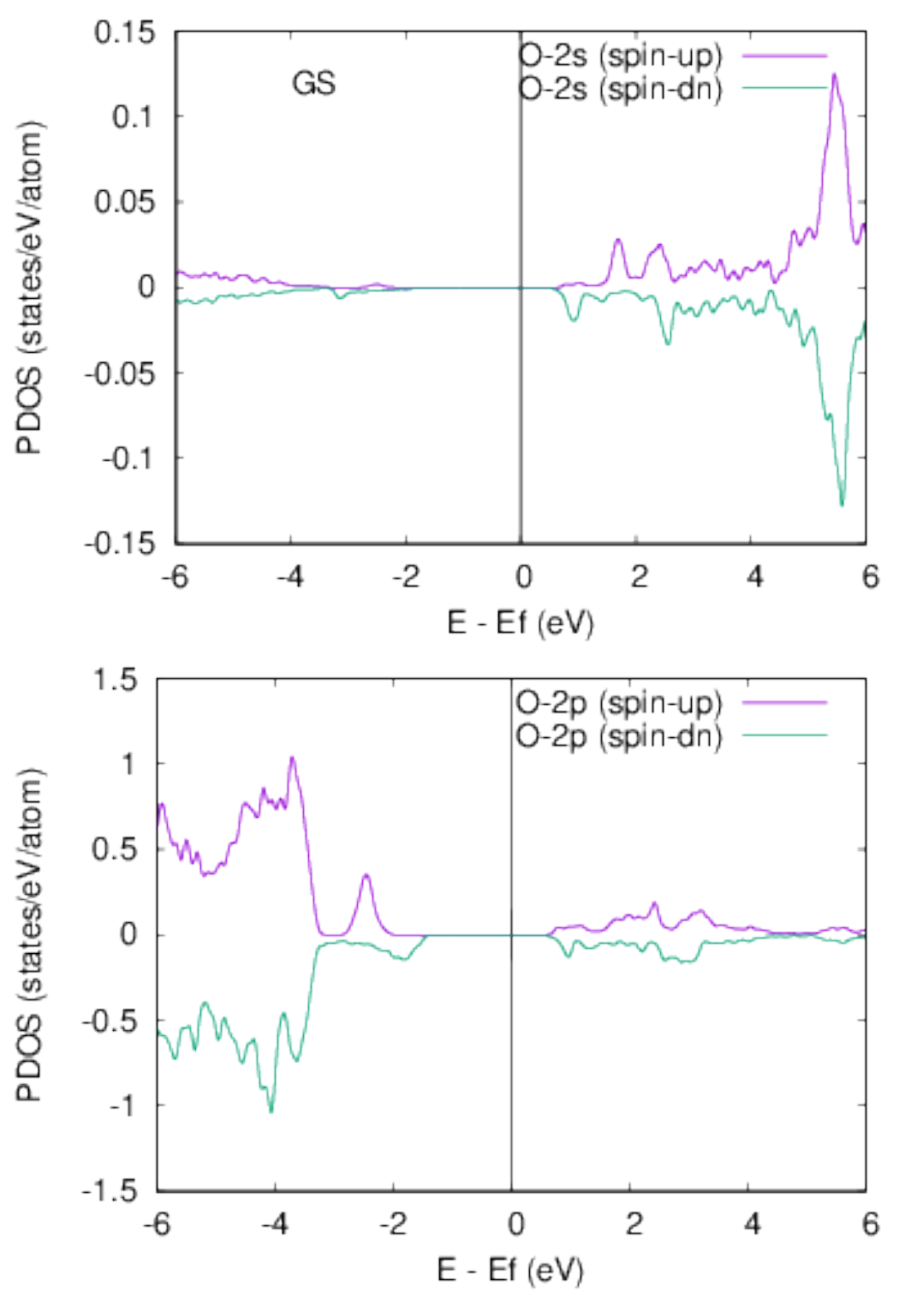}
	\end{subfigure}
	\begin{subfigure}{0.24\textwidth}
		\includegraphics[width=\linewidth]{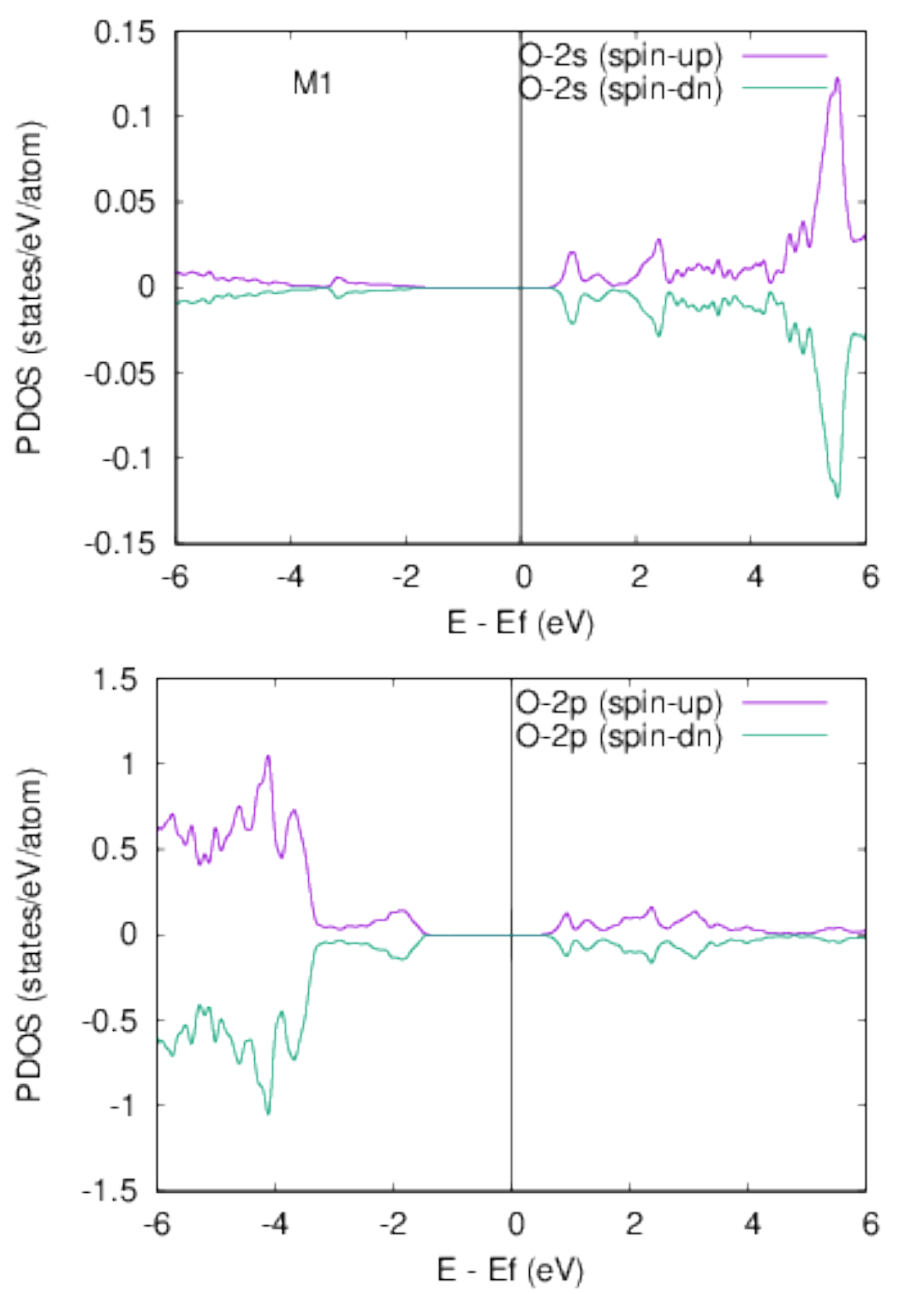}
	\end{subfigure}
	\begin{subfigure}{0.24\textwidth}
		\includegraphics[width=\linewidth]{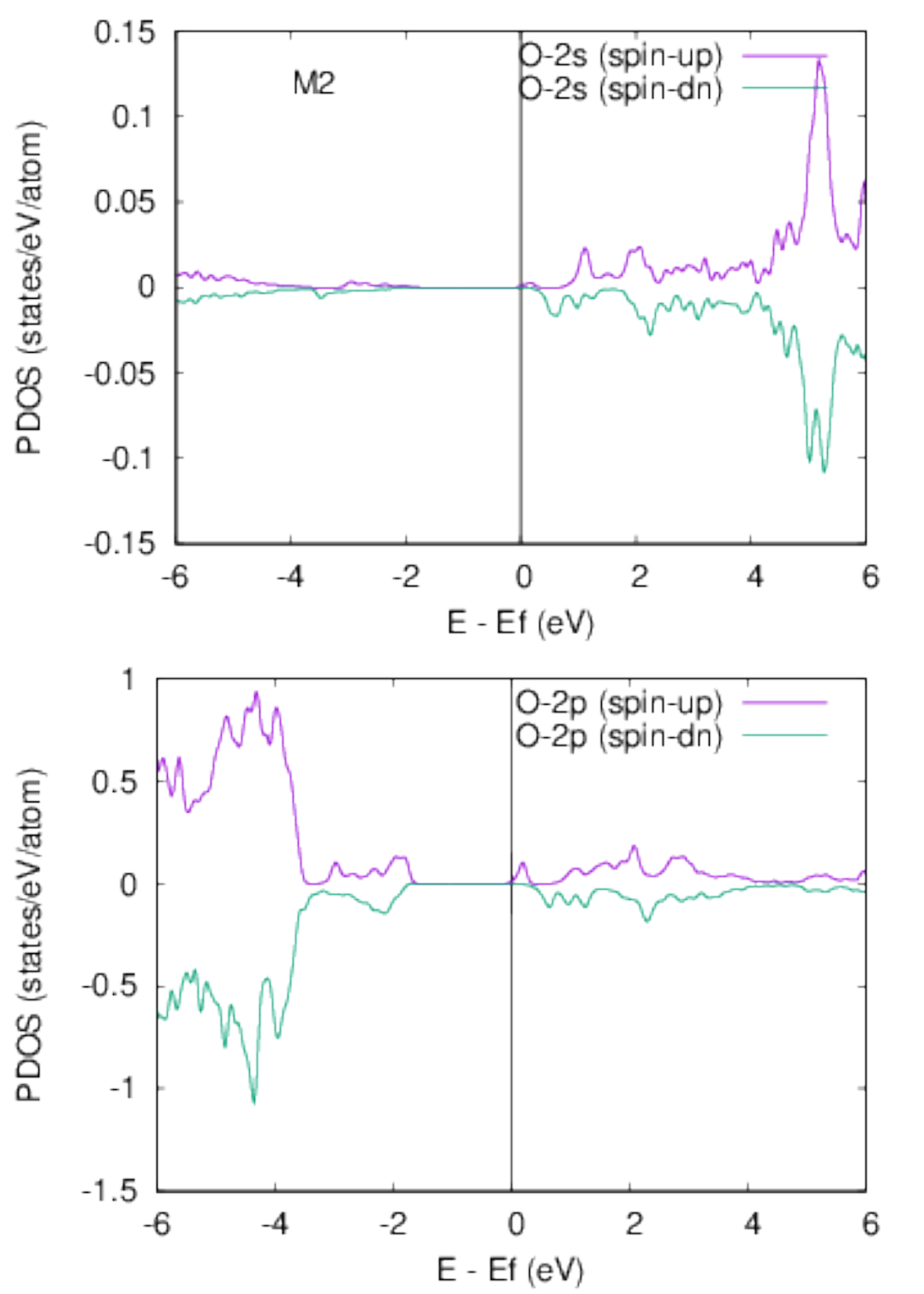}
	\end{subfigure}
	\begin{subfigure}{0.24\textwidth}
		\includegraphics[width=\linewidth]{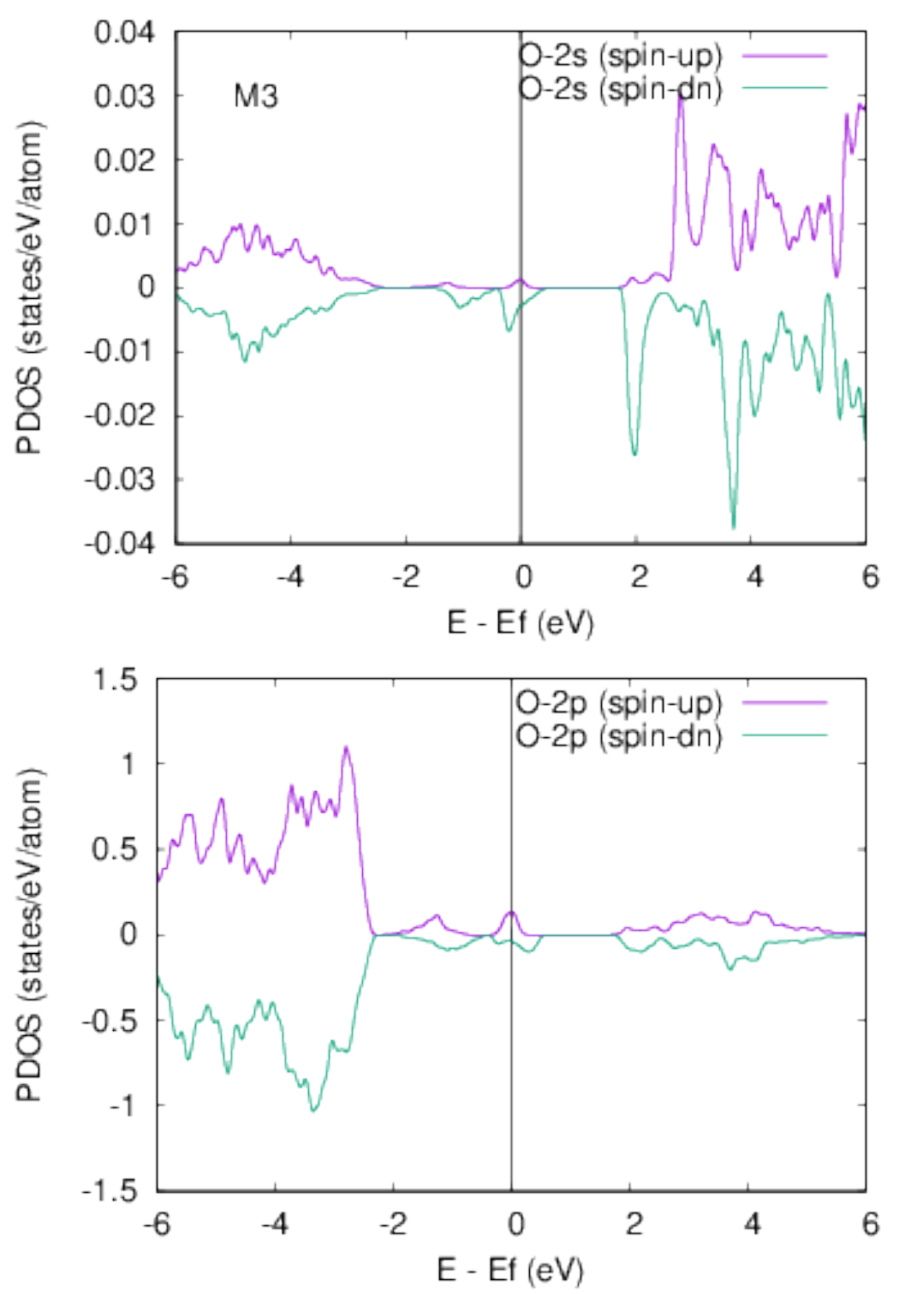}
	\end{subfigure}\\
    \caption{Projected density of states onto two atomic valence atomic orbitals $2s$ and $2p$ of 
    	O-atoms. The $2p$ orbitals have one order of magnitude stronger contributions in DOS than the $2s$ ones of both valence and conduction bands in all GS, 
    	M1, M2, and M3. Here also as in the case of U-atoms, none of the atomic orbitals have contributions to DOS of GS, M1, and M2 states at 
    	the Fermi level, and therefore, these three states show insulating behaviors. However, in M3 both $2s$ and $2p$ atomic orbitals have contributions to DOS 
    	at Fermi level, and so this state has metallic behavior.}
	\label{fig8}
\end{figure*}

In Fig.~\ref{fig8}, we have plotted the contributions of two valence atomic orbitals $2s$ and $2p$ of 
O-atoms for all states GS, M1, M2, and M3. 
As is seen, the $2p$ orbitals have one order of magnitude stronger contributions in DOS than the $2s$ ones of both valence and conduction bands in all GS, 
M1, M2, and M3. Here also as in the case of U-atoms, none of the atomic orbitals have contributions to DOS of GS, M1, and M2 states at 
the Fermi level, and therefore, these three states show insulating behaviors. However, in M3 both $2s$ and $2p$ atomic orbitals have contributions to DOS 
at Fermi level, and so this state has metallic behavior.
Comparing the PDOS's of O-atoms for GS and M1 we see an important difference. As we observed from Fig.~\ref{fig3}, the M1 state is achieved when we used small starting magnetizations around zero; therefore, the spin-up and spin-down contributions are similar which is obtained when we constrain the starting magnetization to zero value. That is, we have a symmetry between the spin-up and spin-down contributions for the O-atoms in M1.

\begin{figure*} 
	\centering
	\begin{subfigure}{.24\textwidth}
		\includegraphics[width=\textwidth]{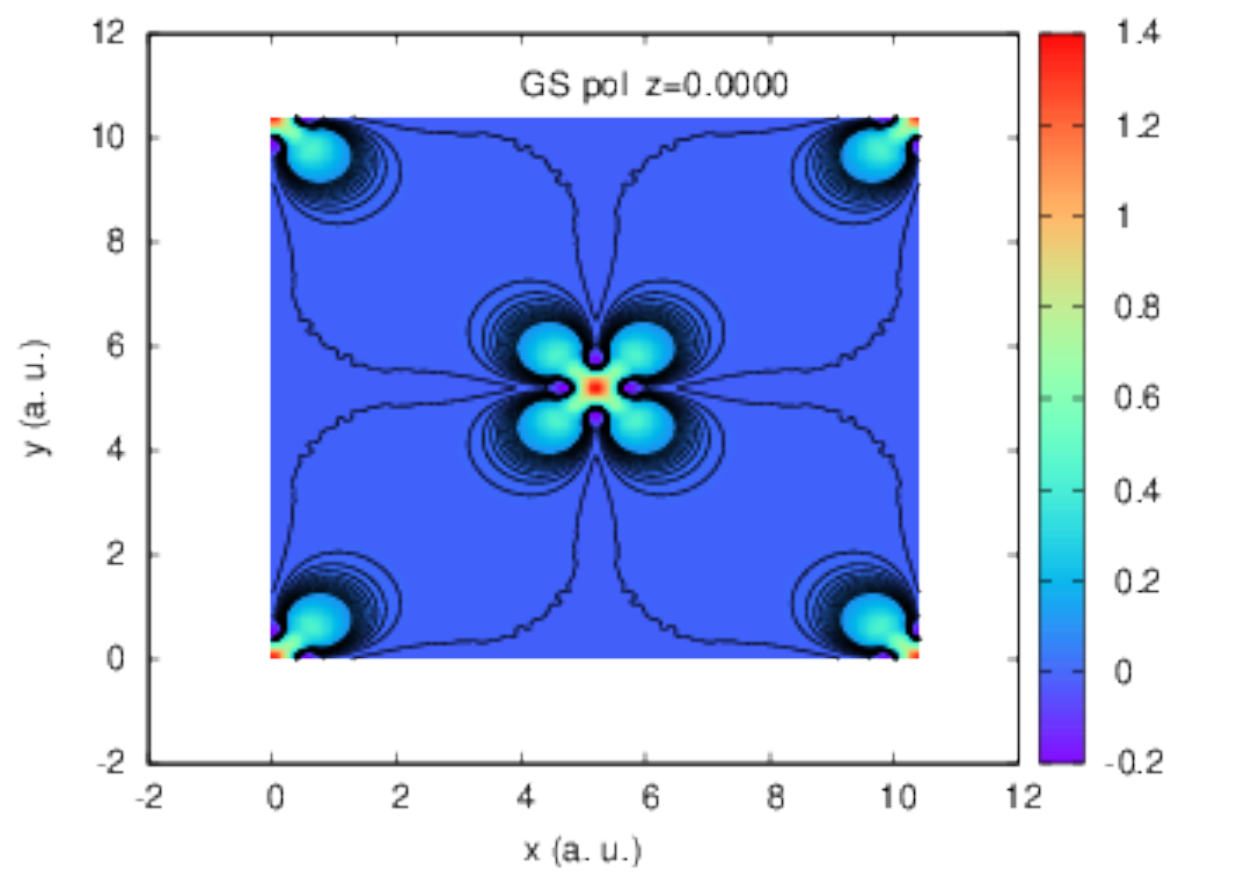}
	\end{subfigure} 
	\begin{subfigure}{.24\textwidth}
		\includegraphics[width=\textwidth]{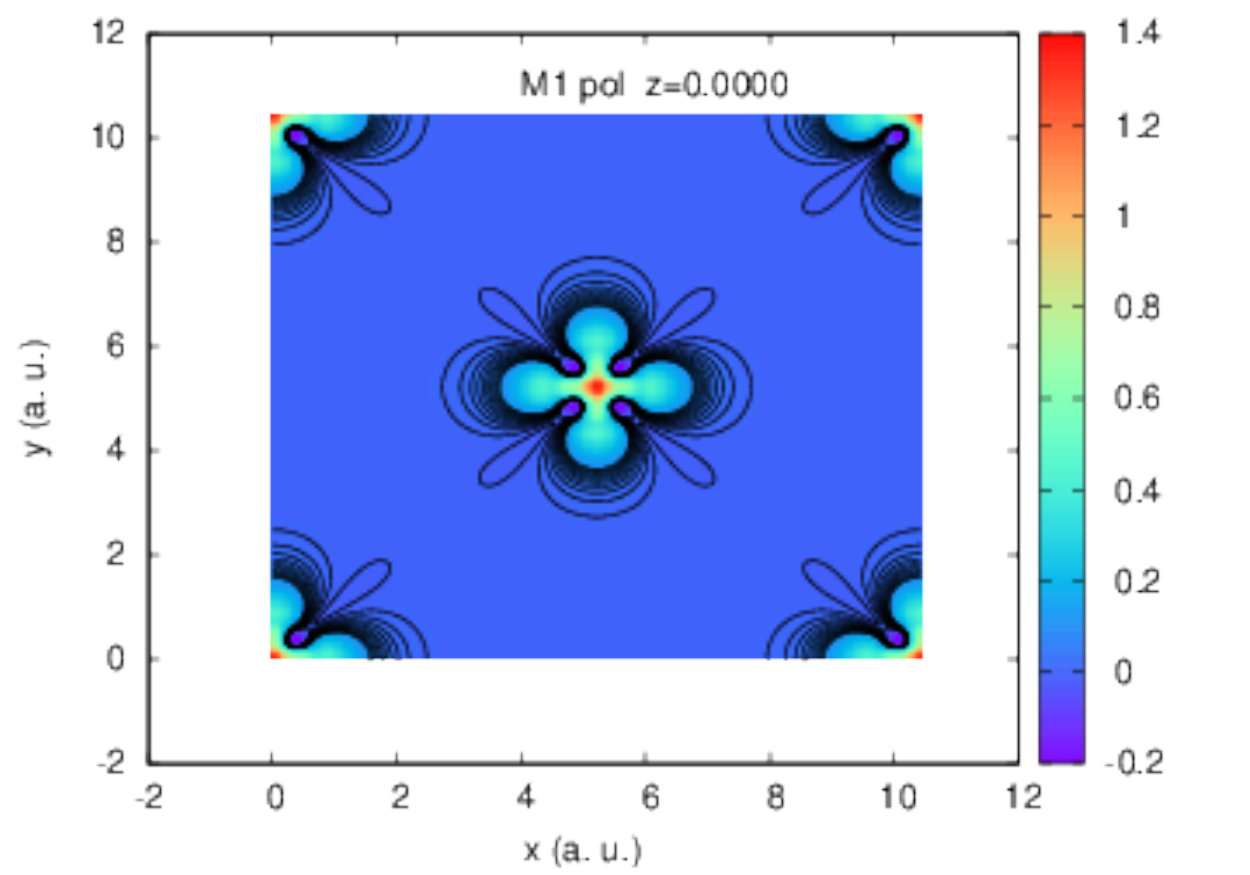}
	\end{subfigure} 
	\begin{subfigure}{.24\textwidth}
		\includegraphics[width=\textwidth]{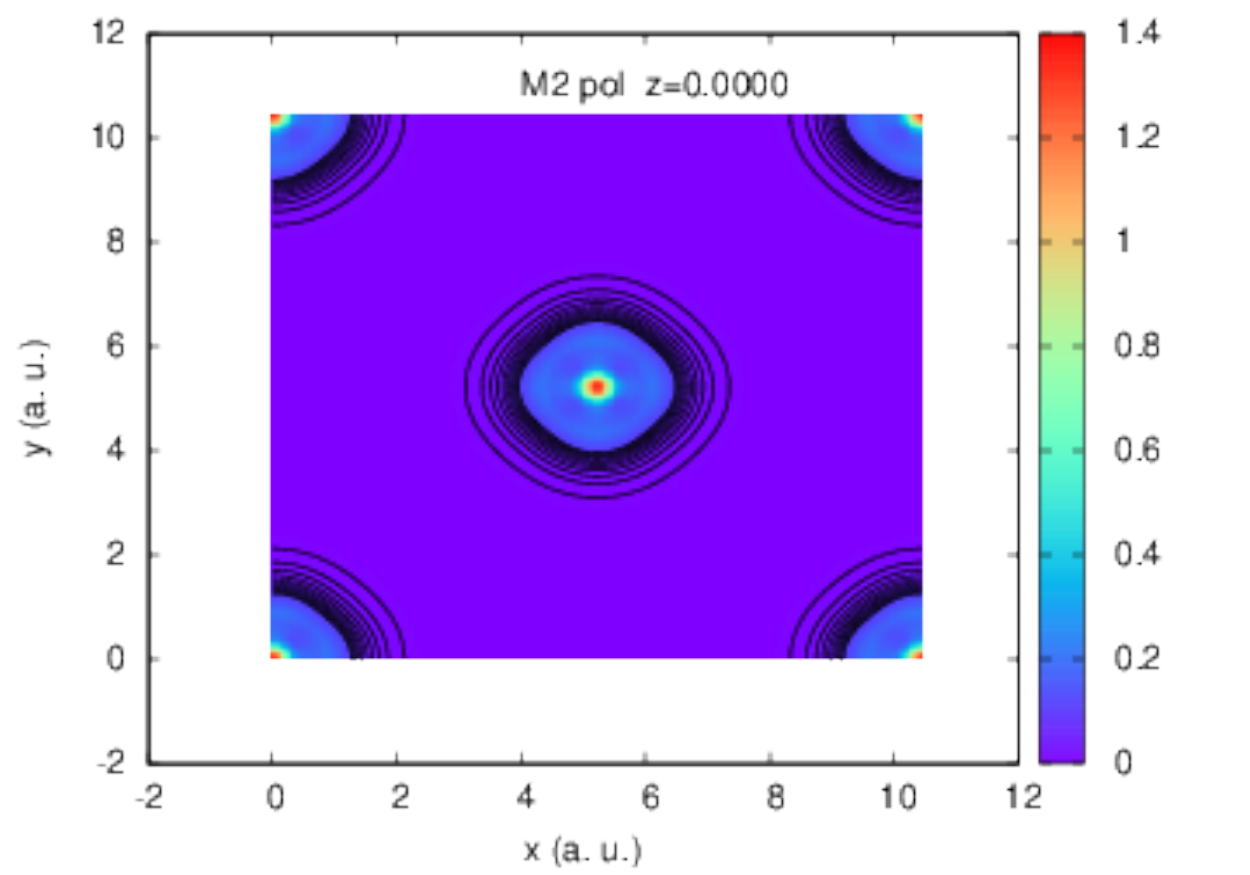}
	\end{subfigure} 
	\begin{subfigure}{.24\textwidth}
	    \includegraphics[width=\textwidth]{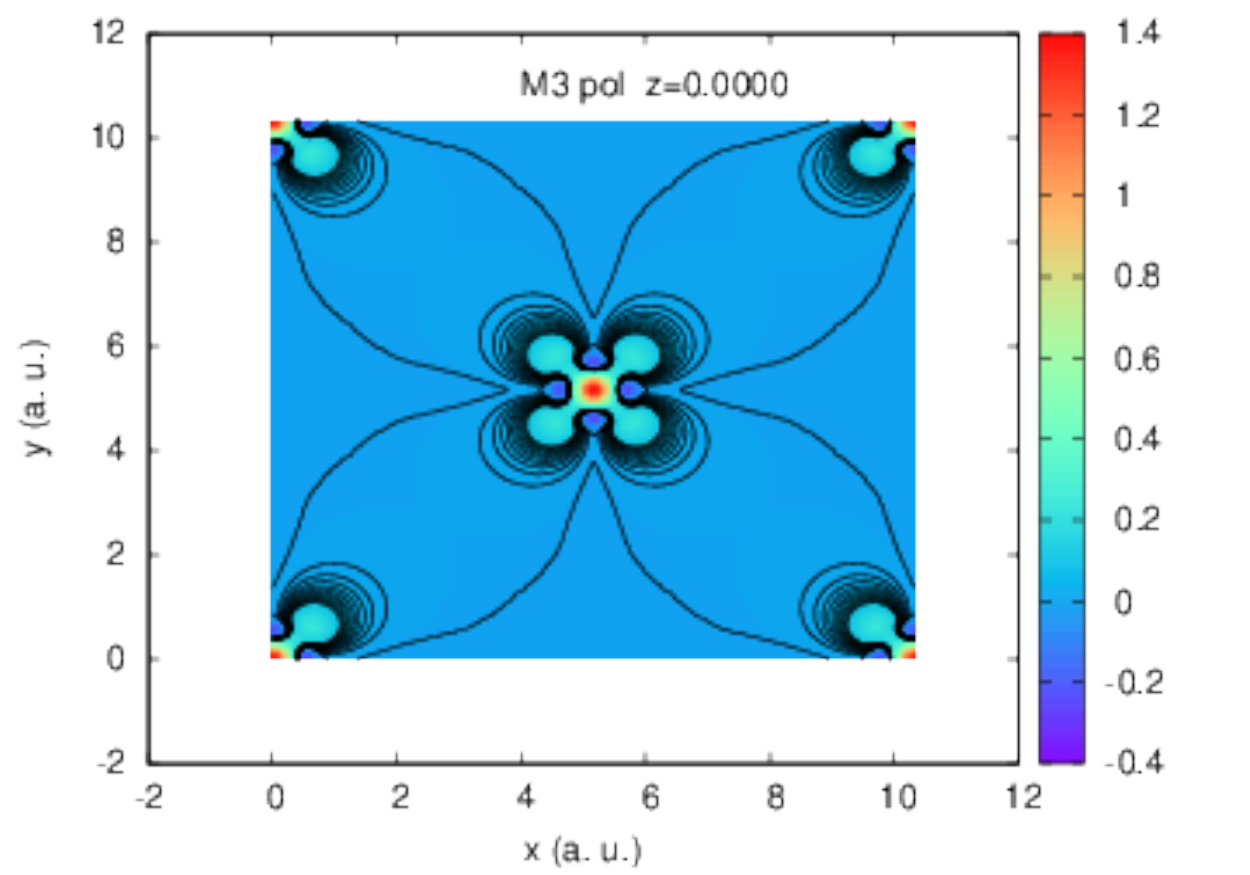}
    \end{subfigure} \\ 
    \begin{subfigure}{.24\textwidth}
		\includegraphics[width=\textwidth]{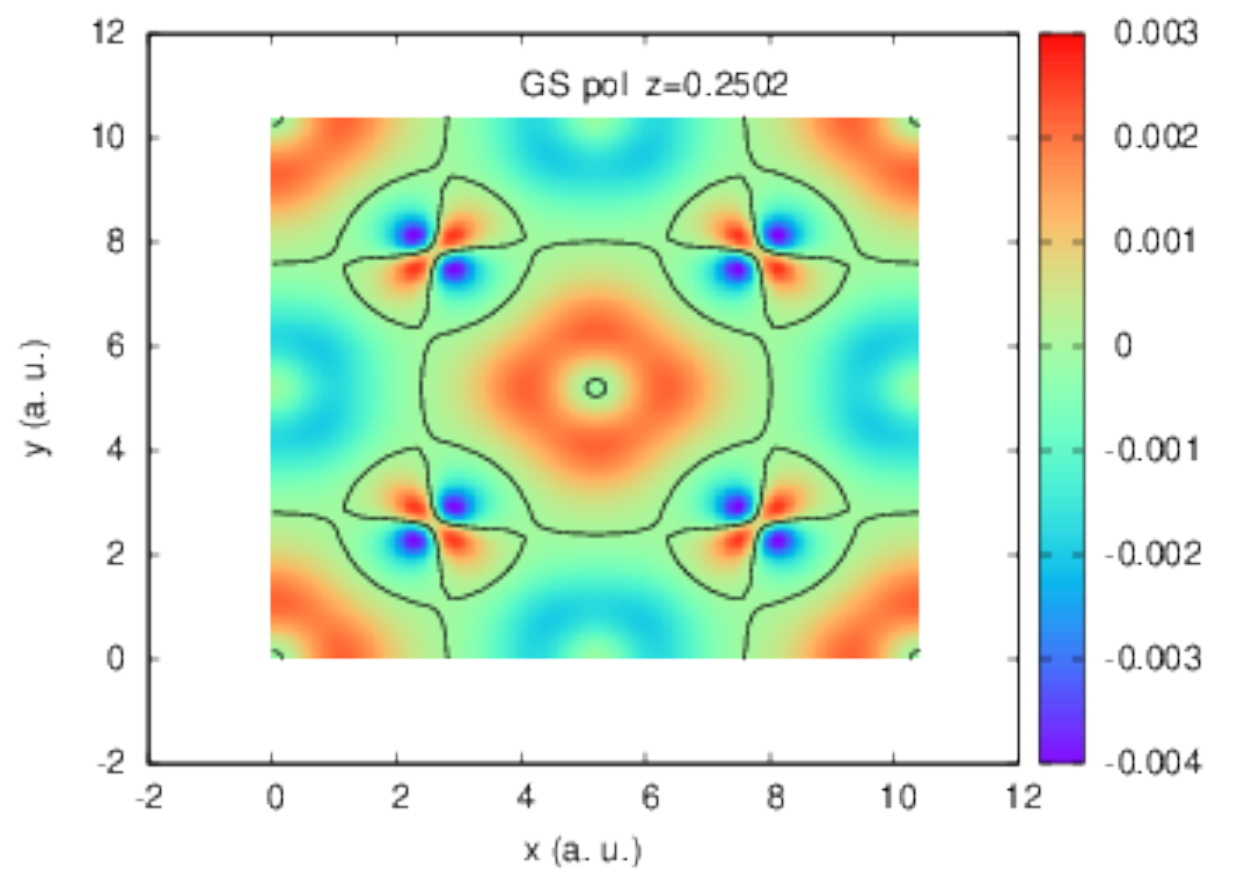}
	\end{subfigure} 
	\begin{subfigure}{.24\textwidth}
		\includegraphics[width=\textwidth]{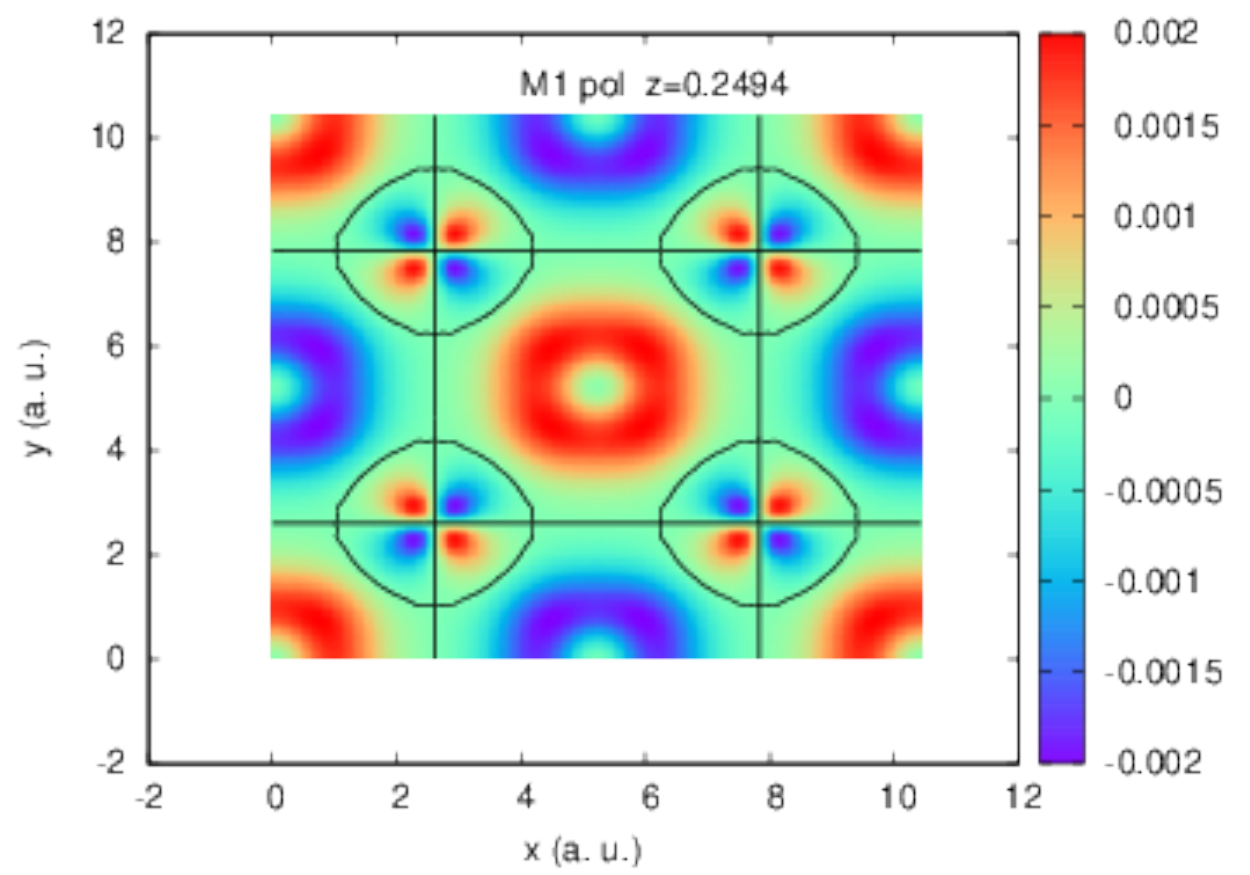}
	\end{subfigure} 
	\begin{subfigure}{.24\textwidth}
		\includegraphics[width=\textwidth]{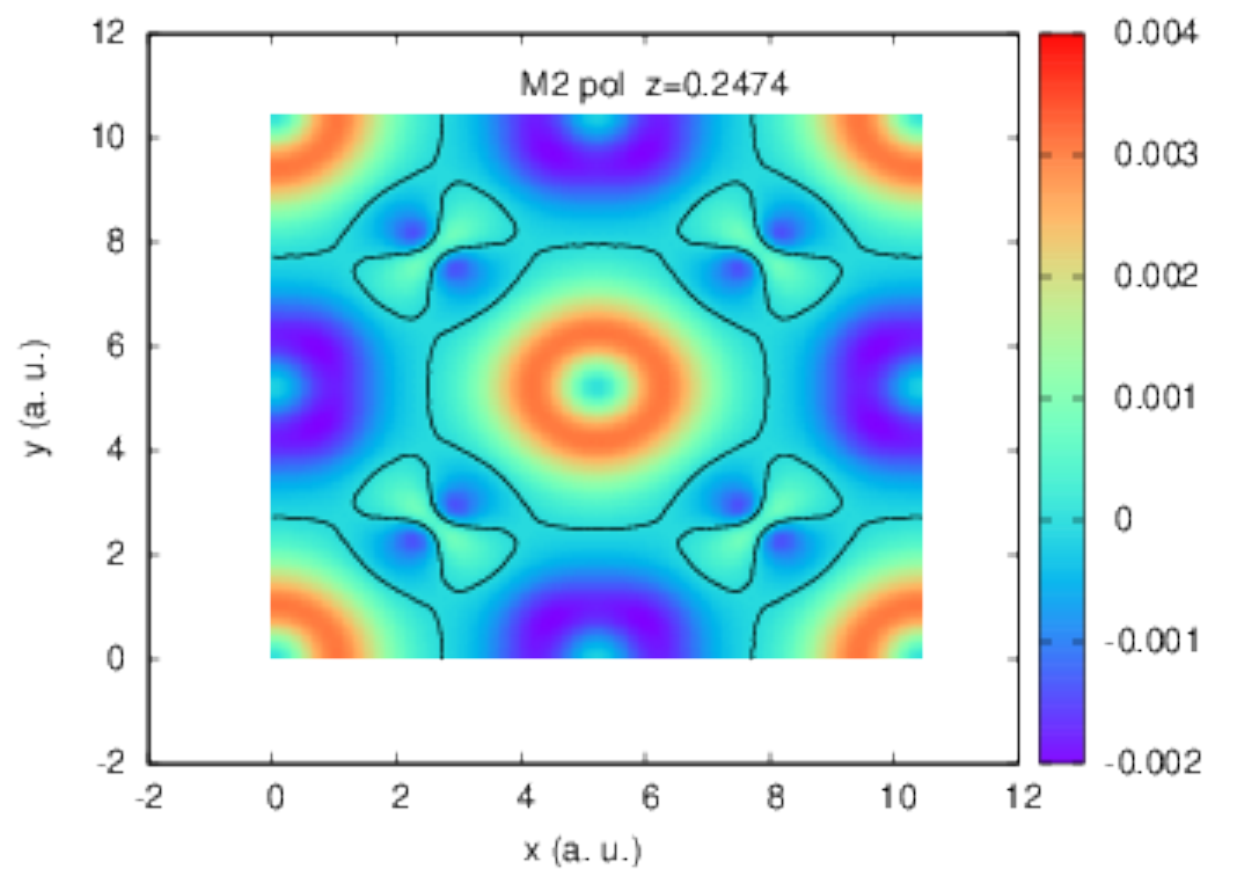}
	\end{subfigure} 
	\begin{subfigure}{.24\textwidth}
		\includegraphics[width=\textwidth]{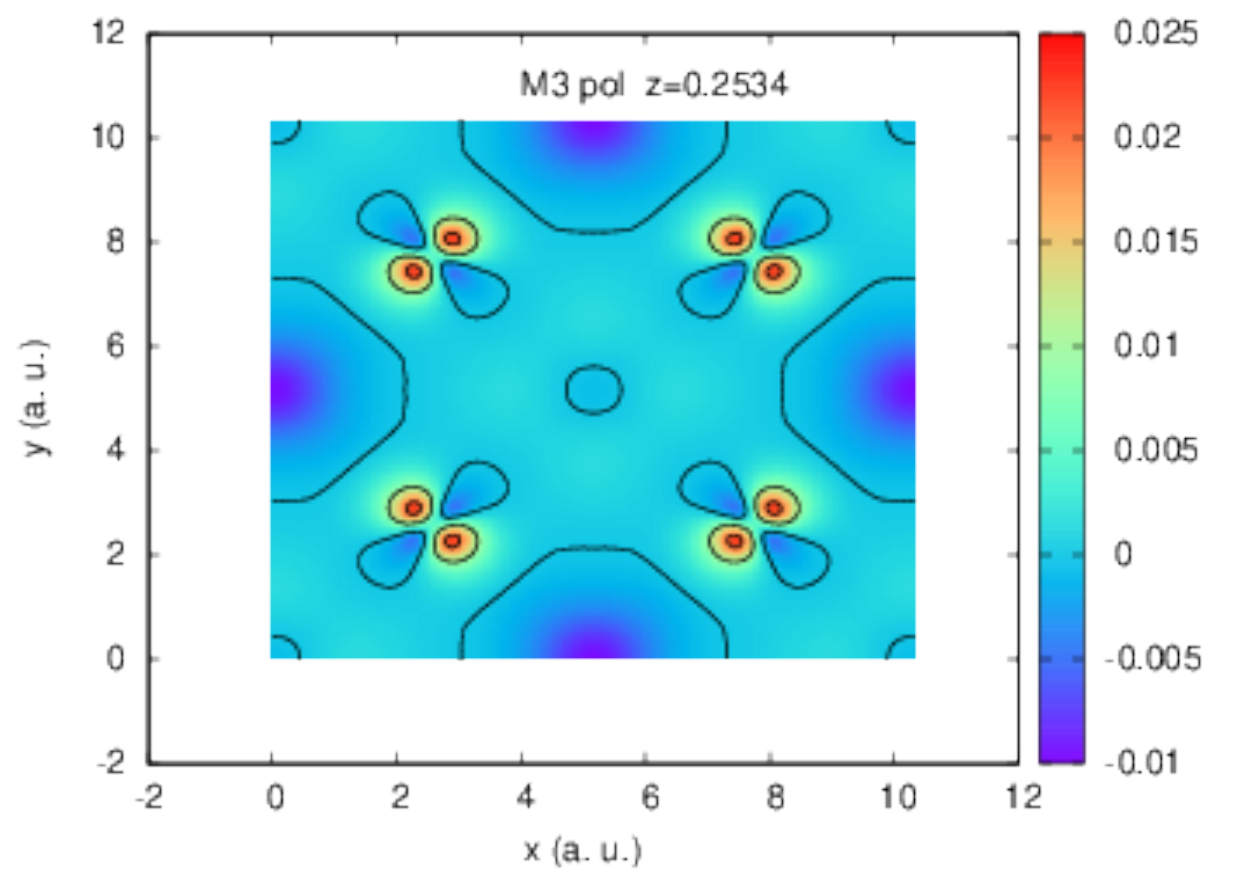}
	\end{subfigure} \\
	\begin{subfigure}{.24\textwidth}
	    \includegraphics[width=\textwidth]{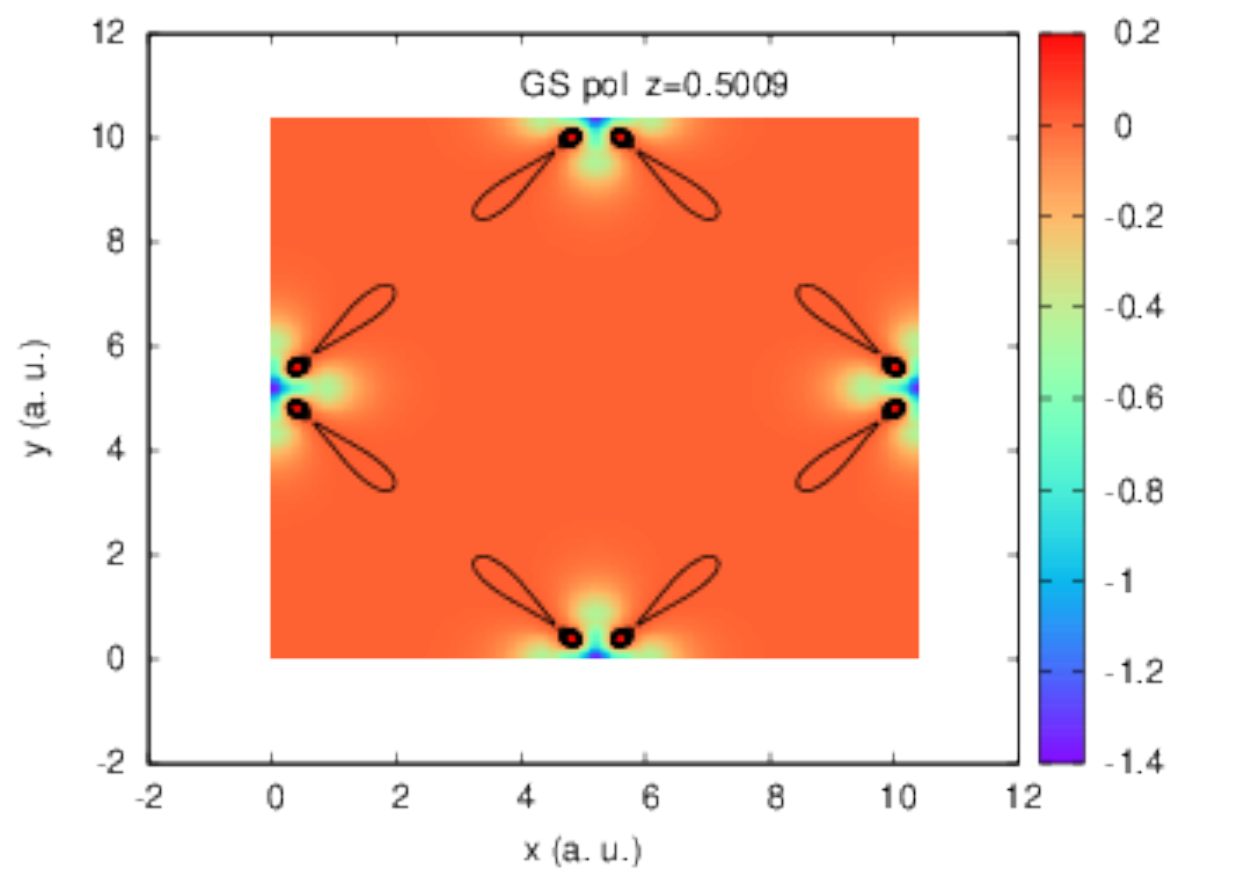}
    \end{subfigure} 
    \begin{subfigure}{.24\textwidth}
	    \includegraphics[width=\textwidth]{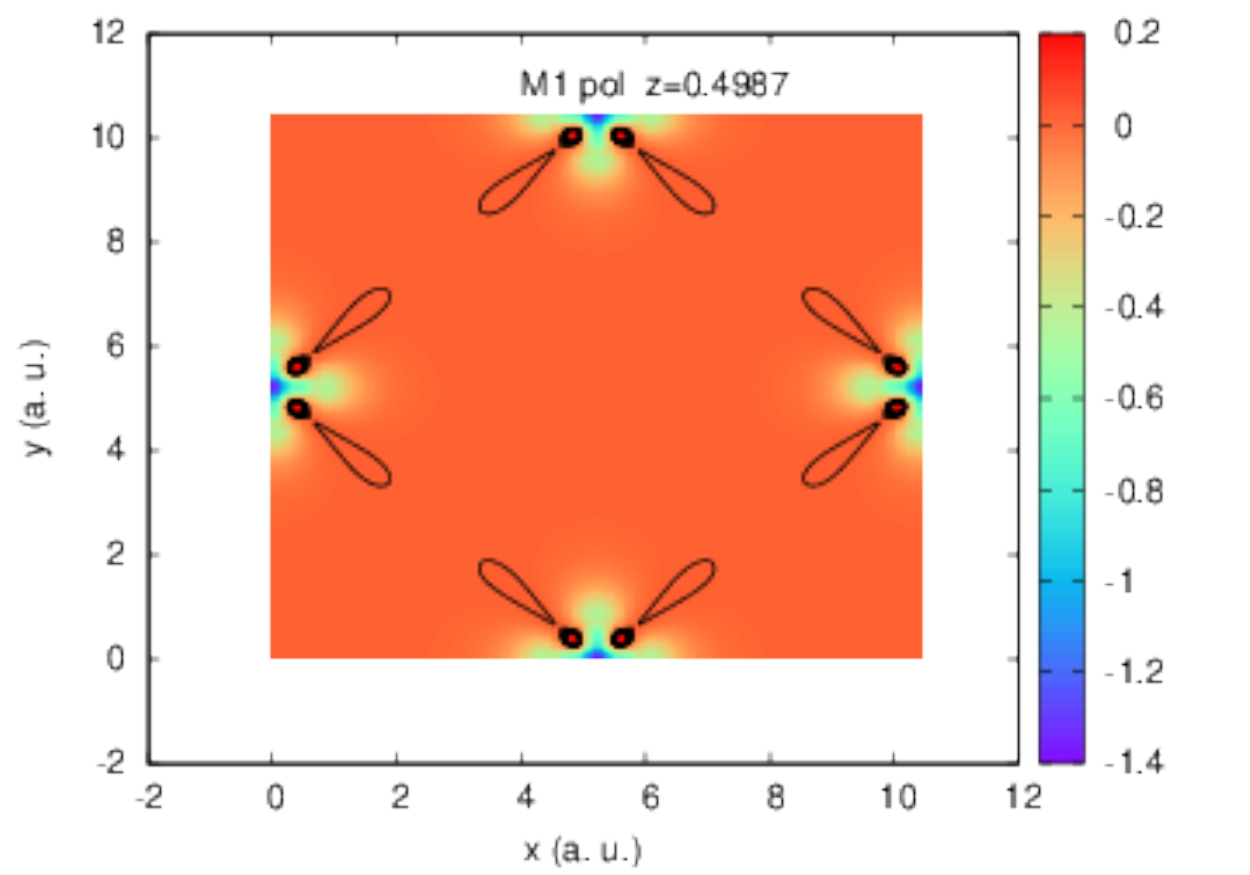}
    \end{subfigure} 
    \begin{subfigure}{.24\textwidth}
	    \includegraphics[width=\textwidth]{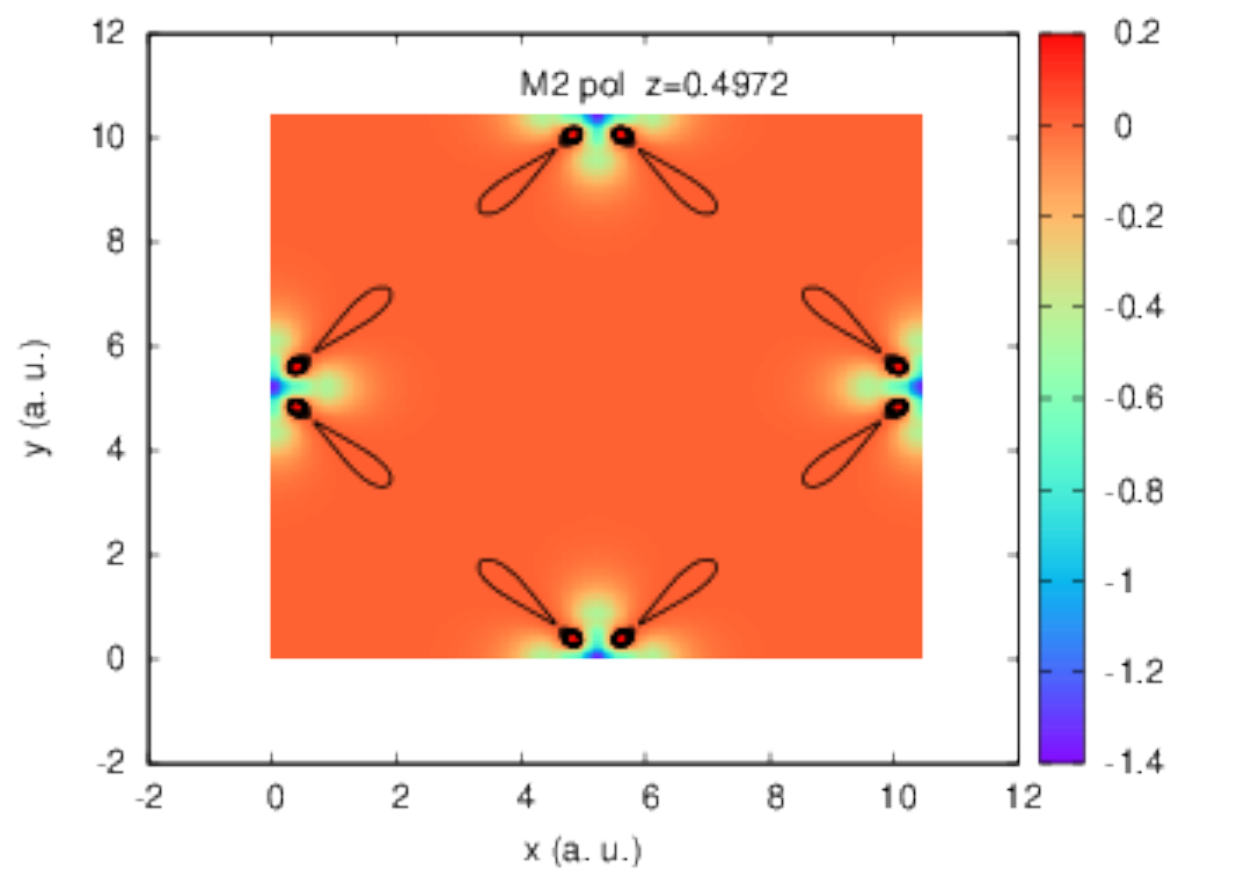}
    \end{subfigure} 
    \begin{subfigure}{.24\textwidth}
	    \includegraphics[width=\textwidth]{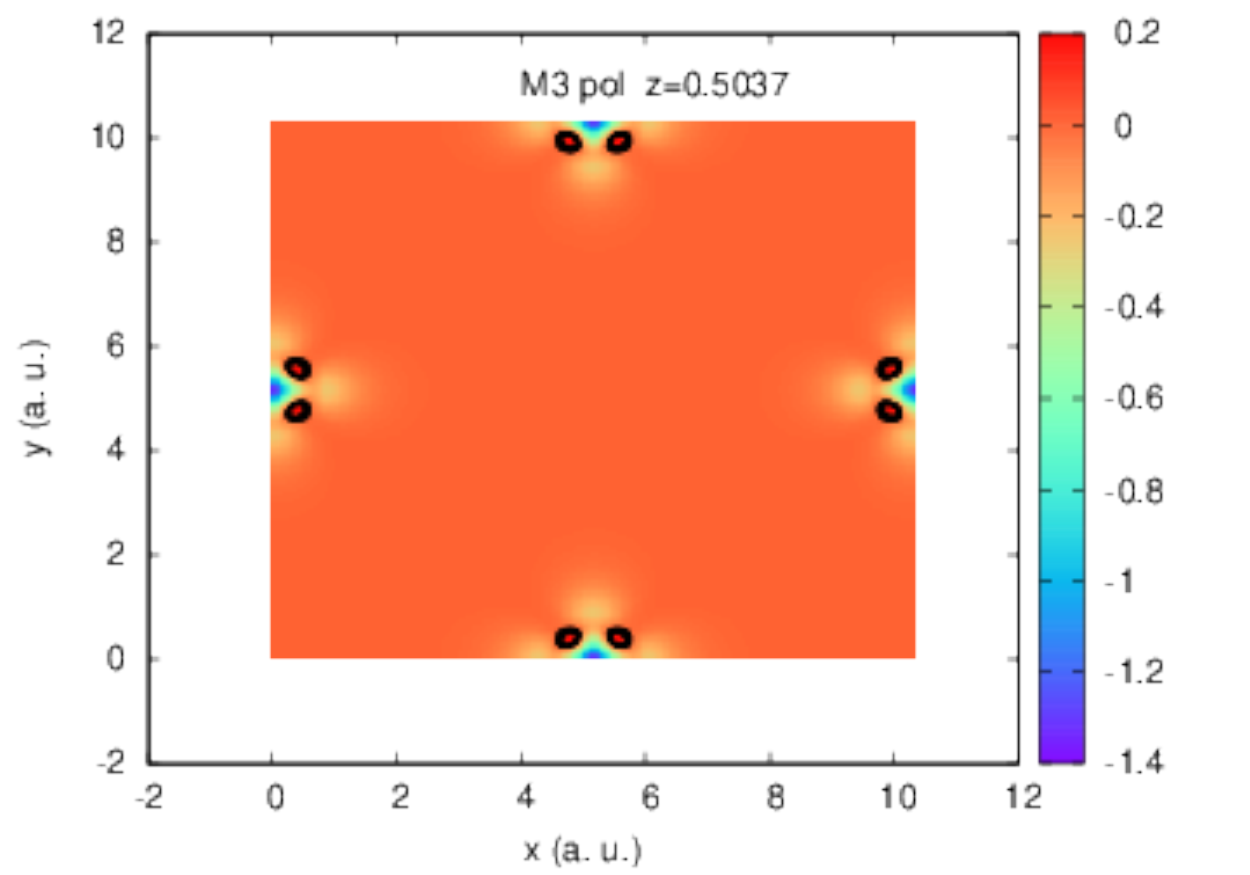}
    \end{subfigure} \\ 
	\caption{Local electronic polarization densities, $(n_\uparrow ({\bf r}) -n_\downarrow ({\bf r}))/(n_\uparrow ({\bf r}) 
		+ n_\downarrow({\bf r}))$ , in atomic units, at equilibrium positions of the planes with $z_1$, $z_2$, and $z_3$ values for GS, M1, M2, and M3 states. As written on sub-figures, the $z$-values are different for the GS, M1, M2, and M3 states. The $z_1$ planes contain the uranium atoms with 
		up-spin configuration in the unit cell. The $z_2$ planes pass through the oxygen atoms of the unit cell, and the $z_3$ planes contain the second type 
		uranium atoms with down-spin
		configurations in the AFM structure. The values of first three 
		states at $z_3$ planes show similar behaviors which are different from that of the M3 state which has metallic behavior. }
	\label{fig9}
\end{figure*}

In Fig.~\ref{fig9}, we have plotted the local electronic polarization densities, $(n_\uparrow ({\bf r}) -n_\downarrow ({\bf r}))/(n_\uparrow ({\bf r}) 
+ n_\downarrow({\bf r}))$ , in atomic units, at equilibrium positions of the planes with $z_1$, $z_2$, and $z_3$ values for GS, M1, M2, and M3 states. From the different equilibrium 
lattice constants for GS, M1, M2, and M3 states, it is clear that these $z$-values are different for the states. The $z_1$ planes contain the uranium atoms with 
up-spin at
$z=0$ in the unit cell. The $z_2$ planes pass through the oxygen atoms of the unit cell, and the $z_3$ planes contain the second type uranium atoms with down-spin
configurations in the AFM structure. The second oxygen layers in the unit cell have the same properties as the first oxygen layers, and so we have not included 
in the plots. It is clearly seen that the polarization values for the $z_1$ and $z_2$ planes are different for all four states, while the values of first three 
states at $z_3$ planes are similar and different from that of the M3 state which has metallic behavior. 

Finally, to check the dependence of the energetic and geometric properties of the GS as well as the meta-stable states on the values of the starting 
magnetizations $\pm 0.5$ for the U1 and U2 uranium atoms in the 1k-order AFM configuration, we have set the starting magnetizations of U1 and U2 atoms to $+1.0$ 
and $-1.0$ respectively, and repeated the simplified all-equivalent oxygen atoms model calculations. The results are summarized in Table~\ref{tab4}.
To distinguish between these states from the previous ones, we have named as GS*, M1*, M2*, M3*, M4*, M5*, and M6*. The first important message of the results
in Table~\ref{tab4} is that the energetic and geometric behaviors of the meta-stable states strongly depend on the initial magnetizations of the U-atoms.
The second important result is that the degeneracies in the meta-stable states were disappeared and all new meta-stable states have different geometric properties 
while the magnetization properties may still have similar properties. The third and final finding is that in all the three studied schemes, we obtain one and 
a unique GS properties.
 
\begin{table}[b]
	\caption{\label{tab4}%
		GS and meta-stable states' properties in the simplified all-equivalent oxygen atom model for the case of $\pm 1.0$ initial magnetizations for U-atoms. The 
		energies are in Ry/(unit formula) and are compared to the GS. Equilibrium lattice constants are in $\AA$, total and absolute magnetizations are in Bohr-magneton/(unit formula). GS*, M1*, M2*, M3*, M4*, M5*, and M6* states are different from each other.}
	\begin{ruledtabular}
		\begin{tabular}{lccccr}
			State &   $\Delta E$    &  $a$ ($c$)      &  Tot. mag.  & Abs. mag. & Occ. \\ \colrule 
			GS*    &   0.00000        &  5.5086 (5.4796)&  $ 0.00$    & 2.165  & 7   \\ 
			M1*    &   0.00140        &  5.5001 (5.4931)&  $ 0.00$    & 2.165  & 2   \\
			M2*    &   0.01184        &  5.5299 (5.4396)&  $ 0.00$    & 2.165  & 6 \\
			M3*    &   0.02115        &  5.5378 (5.4237)&  $ 0.00$    & 2.170  & 1 \\
			M4*    &   0.03008        &  5.5055 (5.4740)&  $ 0.00$    & 2.180  & 1   \\
			M5*    &   0.03008        &  5.5012 (5.4774)&  $ 0.00$    & 2.180  & 2   \\
			M6*    &   0.05879        &  5.4690 (5.5103)&  $-0.04$    & 2.240  & 1 \\			
		\end{tabular}
	\end{ruledtabular} 
\end{table}

\section{Conclusions}\label{sec4}
In this work, it was shown that the "true" GS of 1k-order AFM UO$_2$ system is a spin-symmetry broken state of the electron spin magnetizations of oxygen atoms. In the DFT+U approach for strongly correlated systems, the total energy of the system is a multi-minima function of input parameters 
and one has to be careful to calculate the true ground state properties and avoid assigning one of the meta-stable states as the GS. The occupation-matrix control, simulated-annealing, and U-ramping 
methods, which had been introduced by other researchers, may help one to find lower-energy states but no guarantee to be the lowest-energy state, i. e., the "true" ground state (GS). In this work, a new simple and straight-forward method of 
SMC was introduced which helps to find the "true" GS as well as the meta-stable states 
of 1k-order AFM UO$_2$. It was shown that the GS of this system is achieved when the spin-symmetry of the oxygen atoms was broken. The SMC method was applied in the context of two "all-equivalent oxygen atoms" and "two-inequivalent oxygen atoms" models for the initial magnetizations of the U1 
and U2 atoms set to $+0.5$ and $-0.5$, respectively. In both calculations, the results showed that the GS is obtained for asymmetric values of O-atom starting magnetizations. In the first model, 7 doubly-degenerate states were predicted while in the second model, 17 different states including the GS were predicted.
The GS's in the two models showed the same geometric and energetic properties. The DOS's and PDOS's for different states in the simplified model were compared
and the comparison showed that the GS, M1, and M2 states are insulators while the M3 has metallic behavior. Moreover, it was shown that the $5f$ orbitals of 
the U-atoms have the strongest contribution in the density of states of both valence and conduction bands. To visualize the amount of asymmetry in the 
$n_\uparrow$ and $n_\downarrow$ electron densities for the GS and meta-stable states, we have plotted the electronic polarization densities on the three
different $z$ planes in the unit cell. The plots for the planes containing O-atoms showed different behaviors for all states. The three states of GS, M1, and M2 
showed similar behaviors on the $z_3$ plane.  
To check the sensitivity of the energetic and geometric properties of the GS and meta-stable states on the values of the initial magnetizations  
$\pm 0.5$ for the U1 and U2 uranium atoms in the 1k-order AFM configuration, we have changed these starting magnetizations of U1 and U2 atoms to $+1.0$ 
and $-1.0$ respectively, and recalculated the simplified all-equivalent oxygen atoms model. The results showed 
that the energetic and geometric behaviors of the meta-stable states strongly depend on the initial magnetizations of the U-atoms, and 
the degeneracies in the meta-stable states are also dependent on the choice of initial magnetizations of U-atoms.
The most important result was that the GS properties were unique in the models employed. 
Finally, using the GGA-PBEsol approximation for the XC functional, we have obtained electronic and geometric properties of the GS in excellent agreement
with experimental values.

\section*{Acknowledgement}
This work is part of research program in School of Physics and Accelerators, NSTRI, AEOI.  

\section*{Data availability }
The raw or processed data required to reproduce these results can be shared with anybody interested upon 
sending an email to M. Payami.

\bibliography{revised-payami-SMC_only-arxiv-1400.10.10}

\end{document}